\documentclass[sn-basic]{sn-jnl}

\usepackage{amsmath,amssymb,amsfonts}
\usepackage{algorithmic}
\usepackage{graphicx}
\usepackage{textcomp}
\usepackage{xcolor}
\usepackage[algo2e, ruled]{algorithm2e} 
\usepackage{amsmath,amsfonts}
\usepackage{graphicx}
\usepackage{textcomp}
\usepackage{xcolor}
\usepackage{multirow}
\usepackage{times}
\usepackage{soul}
\usepackage{url}
\usepackage{hyperref}
\usepackage{amsmath}
\usepackage{amsthm}
\usepackage{enumitem}
\usepackage{booktabs}
\usepackage{algorithm}
\usepackage{algorithmic}
\usepackage[switch]{lineno}
\usepackage{subfigure}
\usepackage{makecell}
\usepackage{color}
\usepackage{mathrsfs}
\usepackage[]{collab}

\definecolor{ballblue}{rgb}{0.13, 0.67, 0.8}
\definecolor{yellow}{rgb}{1, 0.3, 0.1}
\collabAuthor{jy}{ballblue}{Jinyang Liu}
\collabAuthor{jz}{magenta}{Jiazhen Gu}
\collabAuthor{gb}{magenta}{Guangba}
\collabAuthor{ww}{yellow}{Wenwei Gu}

\newcommand{\blue}{\textcolor{black}}

\newcommand{\ie}{{\em i.e.},\xspace}
\newcommand{\eg}{{\em e.g.},\xspace}
\newcommand{\name}{KPIRoot+\xspace}

\begin{document}

\title[Article Title]{KPIRoot+: An Efficient Integrated Framework for Anomaly Detection and Root Cause Analysis in Large-Scale Cloud Systems}

\author[1]{Wenwei Gu}\email{wwgu21@cse.cuhk.edu.hk}

\author[1]{Renyi Zhong}\email{ryzhong22@cse.cuhk.edu.hk}

\author[1]{Guangba Yu \footnote{Guangba Yu is the corresponding author.}}\email{guangbayu@cuhk.edu.hk}

\author[4]{Xinying Sun}\email{sunxinying1@huawei.com}

\author[1]{Jinyang Liu}\email{jyliu@cse.cuhk.edu.hk}

\author[3]{Yintong Huo}\email{ythuo@smu.edu.sg}

\author[2]{Zhuangbin Chen}\email{chenzhb36@mail.sysu.edu.cn}

\author[1]{Jianping Zhang}\email{jpzhang@cse.cuhk.edu.hk}

\author[1]{Jiazhen Gu}\email{jiazhengu@cuhk.edu.hk}

\author[4]{Yongqiang Yang}\email{yangyongqiang@huawei.com}

\author[1]{Michael R. Lyu}\email{lyu@cse.cuhk.edu.hk}

\affil[1]{\orgname{The Chinese University of Hong Kong}, \country{Hong Kong SAR}}

\affil[2]{\orgname{Sun Yat-sen University}, \country{China}}

\affil[3]{\orgname{Singapore Management University}, \country{Singapore}}

\affil[4]{\orgname{Huawei Cloud Computing Technology Co., Ltd}, \country{China}}

\abstract{\blue{To ensure the reliability of cloud systems, their runtime status reflecting the service quality is periodically monitored with monitoring metrics, \ie KPIs (key performance indicators). When performance issues happen, \textit{root cause localization} pinpoints the specific KPIs that are responsible for the degradation of overall service quality, facilitating prompt problem diagnosis and resolution. To this end, existing methods generally locate root-cause KPIs by identifying the KPIs that exhibit a similar anomalous trend to the overall service performance. While straightforward, solely relying on the similarity calculation may be ineffective when dealing with cloud systems with complicated interdependent services. Recent deep learning-based methods offer improved performance by modeling these intricate dependencies. However, their high computational demand often hinders their ability to meet the efficiency requirements of industrial applications. Furthermore, their lack of interpretability further restricts their practicality. To overcome these limitations, an effective and efficient root cause localization method, KPIRoot, is proposed. It integrates both advantages of similarity analysis and causality analysis, where similarity measures the trend alignment of KPI and causality measures the sequential order of variation of KPI. Furthermore, it leverages symbolic aggregate approximation to produce a more compact representation for each KPI, enhancing the overall analysis efficiency of the approach. However, during the deployment of KPIRoot in cloud systems of a large-scale cloud system vendor, Cloud $\mathcal{H}$. We identified two additional drawbacks of KPIRoot: 1. The threshold-based anomaly detection method is insufficient for capturing all types of performance anomalies; 2. The SAX representation cannot capture intricate variation trends, which causes suboptimal root cause localization results. We propose \name to address the above drawbacks. The experimental results show that \name outperforms eight state-of-the-art baselines by 2.9\%$\sim$35.7\%, while time cost is reduced by 34.7\%. Moreover, we share our experience of deploying KPIRoot in the production environment of a large-scale cloud provider Cloud $\mathcal{H}$\footnote{Due to the company policy, we anonymize the name as Cloud $\mathcal{H}$.}.}}


\maketitle

\section{Introduction}

Large-scale cloud systems have become the backbone of modern computing infrastructure, offering unprecedented scalability and flexibility. Cloud platforms such as Microsoft Azure, Amazon Web Services, and Google Cloud Platform provide cost-effective services to users worldwide on a $7 \times 24$ basis~\cite{lin2018predicting,yu2024deep}. However, the inherent complexity and scale of these systems inevitably lead to performance issues, including slow application response times, network latency spikes, and resource contention~\cite{lin2016log,kuang2024knowledge}. These issues can result in violations of Service Level Agreements (SLAs), causing user dissatisfaction and financial losses for both service providers and consumers~\cite{liu2023incident}. Consequently, the prompt identification and resolution of performance issues have become critical concerns for cloud vendors and users alike~\cite{liu2016using}. Addressing these challenges is essential for maintaining the reliability and efficiency of cloud services in an increasingly digital world.

Cloud vendors usually collect real-time key performance indicators (KPIs) to monitor the health status of their services~\cite{su2019coflux}. Anomaly detection is conducted over these KPIs to identify performance issues based on this KPI data~\cite{zhao2019label, zhao2020real, huang2022transferable}. For example, if the resource utilization rate is continuously high, it may indicate an imminent service overload and performance degradation. However, due to the scale of cloud systems, it is infeasible to analyze the KPI of each instance (\eg VM and container) individually. Since a cloud service typically consists of many instances, a common way is to monitor specific KPIs that can reflect the overall performance of the service, \eg latency, error count, and traffic, which we refer to as \emph{alarm KPIs}. Automated performance issue detection can thus be realized through configuring alerting rules or performing anomaly detection algorithms on such alarm KPIs. These underlying KPIs of individual instances or VMs within a cloud service may not be directly analyzed due to the scale of cloud systems. However, their collective behavior significantly influences the alarm KPIs. 

When a performance issue is detected (\ie the alarm KPI is abnormal), it is crucial to identify the root cause~\cite{soldani2022anomaly} (\eg which underlying instances cause the abnormal performance of the service). However, pinpointing the root cause is a non-trivial task since the monitored alarm KPI is highly aggregated and often derived~\cite{yan2022cmmd}, \ie the correlation between the underlying KPIs and the alarm KPI is complicated and hard to understand. Even experienced software reliability engineers (SREs) can struggle to pinpoint the specific KPIs that contribute to the root cause. Such a manual approach is like finding a needle in a haystack, which is tedious and time-consuming. Hence, the automated root cause localization method is an urgent requirement for prompt performance issue resolution.

In particular, a practical root cause localization approach for KPIs from cloud systems should meet the \emph{efficiency} and \emph{interpretability} requirements~\cite{yu2023nezha}. Specifically, due to the huge volume of underlying KPIs and the tight time-to-resolve pressure, the approach needs to be able to process large amounts of data (\eg thousands of KPIs) efficiently (\eg in seconds). Furthermore, the approach should produce interpretable results to help engineers take effective remedy actions, which is essential in the maintenance of cloud systems. Existing root cause localization methods typically adopt statistics or deep learning models. Statistic-based methods adopt Kendall, Spearman, and Pearson correlation to compute the linear relationships between KPIs and find the root cause~\cite{yin2016cloudscout}. However, these methods have high computational costs to calculate the correlation for every KPI pair and also suffer from low accuracy in handling complicated KPIs from cloud systems~\cite{yang2021aid}. Some recent studies~\cite{yan2022cmmd} adopt deep learning models (\eg graph neural networks) to model the KPI relationships for root cause localization. However, such methods suffer from high computation costs and lack interpretability~\cite{zhang2022improving, zhang2024curvature}.

\blue{To address the above limitations, a root cause localization framework, KPIRoot~\cite{gu2024kpiroot}, is proposed to identify the root cause underlying KPIs when an anomaly in the monitored alarm KPI is detected in cloud systems. To meet the efficiency requirement, KPIRoot first adopts the Symbolic Aggregate Approximation (SAX) representation to downsample the time series data of KPIs and facilitate extracting the anomaly segments. By filtering out the normal KPI data, KPIRoot can focus on anomaly patterns instead of the whole time series, which optimizes efficiency. Then, KPIRoot conducts both the similarity and causality analysis to localize the root cause KPIs. Specifically, underlying KPIs with a high similarity of anomaly patterns to the alarm KPI are more likely to trigger the alert and be the root causes. On the other hand, causality analysis is used to validate the cause and effect in the temporal dimension, \ie the anomaly pattern of root cause KPIs should happen before that of the alarm KPI. Finally, KPIRoot combines the similarity and causality analysis results to produce a correlation score for each underlying KPI. The higher the score, the more likely the KPI is the root cause. The time complexity of KPIRoot is $\mathcal{O}(\sqrt{n})$ ($n$ is the length of the KPIs), which allows it to process thousands of KPIs in seconds, thus facilitating the resolution of real-time performance issues. }

\blue{However, we identified several drawbacks of KPIRoot. Firstly, the threshold-based anomaly detection method employed by KPIRoot, while effective in identifying trend anomalies, struggles to detect seasonal and point anomalies. This limitation is particularly highlighted in performance issues reflected in KPIs, where seasonal fluctuations and sudden spikes or drops are common and critical to accurate anomaly detection. Secondly, although the SAX representation utilized in KPIRoot enhances the efficiency of root cause localization by downsampling the KPIs, it may not effectively capture intricate variation trends. This limitation arises from its reliance on segment averages, which can obscure variation trend details in the data essential for accurate root cause localization.}

\blue{This paper extends our preliminary work, which appears as a research paper of ISSRE 2024~\cite{gu2024kpiroot}. In particular, we extend our preliminary work in the following directions:}

\begin{itemize}[leftmargin=*]
    \item \blue{We propose \name, an extended version of the KPIRoot framework introduced in our preliminary work~\cite{gu2024kpiroot}. There are two major differences in \name compared to KPIRoot. Firstly, anomaly detection is positioned as a critical precursor to root cause localization. We reveal that the original KPIRoot framework struggles to detect all types of anomalies, which are pivotal for accurate root cause localization in some cases. To address this deficiency, we have implemented a time series decomposition-based method. By supplementing the original approach based on trend variation with time series decomposition and a U-Net autoencoder, \name significantly improves the accuracy of anomaly detection, thus improving the subsequent root cause analysis phase. Secondly, the original Symbolic Aggregate Approximation (SAX) representation employed in KPIRoot falls short of effectively capturing intricate trends and variations due to its dependence on segment averages. This can obscure critical behavioral patterns. To overcome these limitations, \name incorporates an Improved SAX representation (ISAX) that further incorporates trend variation indicators. Our experiments show that \name performs better in terms of root cause localization accuracy but requires a similar execution time when compared with KPIRoot.}
    \item \blue{We conduct a comprehensive evaluation of anomaly detection performance across different models, an aspect that was overlooked in KPIRoot.}
    \item \blue{We strengthen our experimental part by including NDCG in our evaluation metrics, specifically NDCG@5 and NDCG@10. This metric measures how easily engineers can find the culprit VMs, which is crucial in our scenarios as the most relevant root causes are prioritized for investigation.}
    \item \blue{We conduct a sensitivity analysis on the parameters used in \name. The results demonstrate that our approach maintains robustness within a reasonable interval of parameter values.}
    \item \blue{During the deployment of KPIRoot in our Cloud $\mathcal{H}$, we identified several failure cases that highlighted its limitations. We share our industrial experiences and insights on how \name addresses these issues.}
\end{itemize}

To evaluate the effectiveness of our proposed \name, we conducted extensive experiments based on large-scale real-world KPI data from a large cloud vendor. The experimental results demonstrate that \name can pinpoint root cause KPIs more accurately compared with seven baselines with an F1-score of 0.882 and Hit Rate@10 of 0.946. On the other hand, \name largely reduces the computation cost with an execution time of around 8 seconds, which facilitates engineers diagnosing root causes in real time. In particular, we have successfully deployed our approach in the cloud service system of Cloud $\mathcal{H}$ since \textit{Aug 2023} and successfully localized the true root cause of ten performance issues of emergence level with 100 accuracies without affecting the customer. We also share industrial experience in practice.

\blue{We summarize the main contributions of this work, which form a super-set of those in our preliminary study, as follows:}

\begin{itemize}[leftmargin=*]
    \item \blue{We introduce \name, an effective and efficient method to localize the underlying KPIs that cause the anomaly, which is an improved version of KPIRoot. \name adopts the Improved SAX representation for downsampling and combines both the similarity and causality of anomaly patterns of KPIs to identify the root cause. Such designs meet the practical requirements of efficiency and interpretability, making KPIRoot feasible to deploy in large-scale cloud systems. We further strengthen the anomaly detection part to make it effective for different anomaly types.}
    \item Extensive experiments on three industrial datasets collected from Cloud $\mathcal{H}$'s large-scale cloud system demonstrate the effectiveness of \name, \ie 0.882 F1-score and 0.946 Hit@10 rate. The average execution time of \name is around 8 seconds, significantly outperforming seven state-of-the-art baselines. 
    \item We have successfully deployed \name into the troubleshooting system of a large-scale cloud service system of Cloud $\mathcal{H}$ since \textit{Nov 2022}. It has successfully analyzed ten emerging performance issues with 100 accuracies, and none of the issues affected the customer. The success stories of our deployment confirm the applicability and effectiveness of our method.
\end{itemize}

\section{Background and Motivation}

In this section, we present a comprehensive overview of KPI-based root cause analysis in cloud service systems and demonstrate its practical application through a case study of root cause localization in Cloud $\mathcal{H}$, a large-scale production cloud environment.

\begin{figure*}
\centering
\includegraphics[width=.9\linewidth]{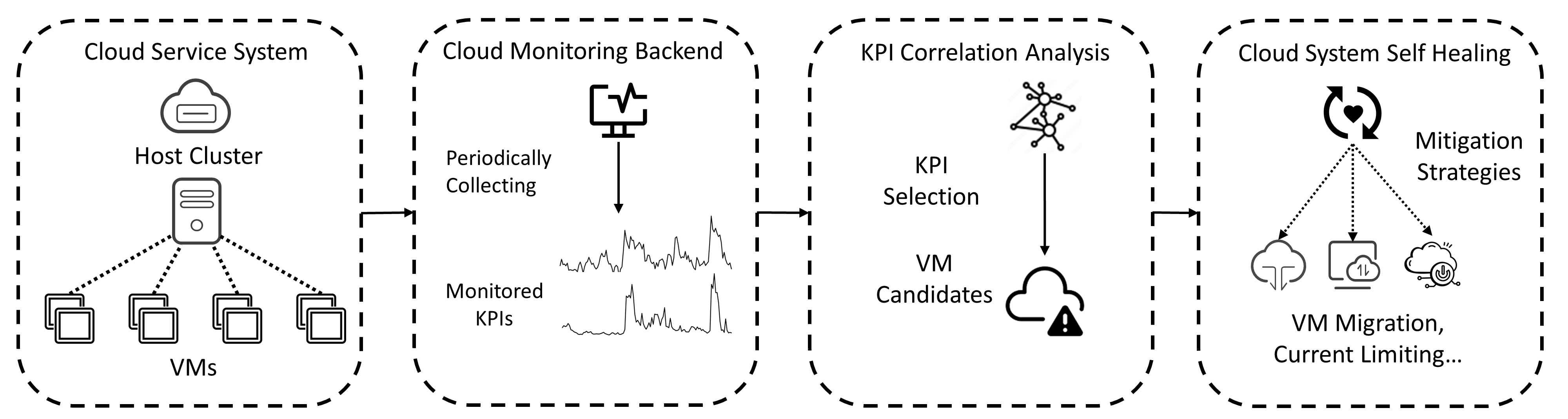}
\vspace{-4pt}
\caption{The Overall Pipeline of Root Cause Localization in Cloud $\mathcal{H}$}
\vspace{-6pt}
\label{framework}
\end{figure*}

\subsection{KPI-based Root Cause Localization in Cloud Systems}

Ensuring performance and reliability in cloud systems is of great importance. Performance anomalies like hardware malfunctions, network overloads, and security violations can significantly influence the performance of cloud systems and violate SLA~\cite{sharma2023sla}. Consequently, the need for run-time status and performance monitoring of cloud systems is in demand. Key Performance Indicators (KPIs) serve as informative tools that monitor the overall status of various components of cloud systems~\cite{cheng2023ai}, providing helpful insights that aid in the identification of potential anomalies~\cite{singh2023predictive}, and even proactively predicting these performance issues before they escalate into catastrophic failures~\cite{tuli2021start}. Some common KPIs in cloud systems include CPU usage, memory usage, network bandwidth, latency, error rates, and service QPS (queries per second).

The cloud service system has become increasingly huge in scale and produces larger volumes of monitoring data. The highly interconnected nature of cloud systems causes problems, such as performance failures, which can spread from one component to another component. Consequently, the failure diagnosis, root cause localization, and performance debugging in large cloud systems are more complex than before~\cite{wang2019grano, qiu2020causality}. In real-world applications, monitoring a large number of KPIs is computationally intensive. Thus, a more practical way is to monitor the aggregated KPI and configure alerts.

Specifically, in large-scale cloud service clusters, large amounts of virtual machines (VMs) operate concurrently to provide tenants with various services. A special KPI is the ``\textit{alarm KPI}'' that triggers alerts when a performance issue like an overload of CPU usage in the entire cluster happens. In large-scale cloud systems, service may consist of large amounts of VMs working together to respond to cloud users' demands~\cite{wickremasinghe2010cloudanalyst}. Given the scale of these systems, individual monitoring of each VM becomes infeasible. Instead, software reliability engineers often utilize alarm KPIs as a more effective approach to oversee the overall performance of the service. When the alarm KPI indicates abnormal activity, it becomes crucial to identify which VMs are the root causes. The root cause refers to the specific VMs that trigger the anomaly within the alarm KPI. For instance, if the alarm KPI is triggered due to a fairly high CPU usage, the root cause could be the particular VMs that directly cause the resource overload. Such a setup allows for the proactive identification of performance issues. In addition to the alarm KPI, other KPIs monitor the bytes per second (bps) and packets per second (pps) of each VM in the cluster~\cite{latah2019artificial}. These KPIs offer valuable insights into the data traffic of each user, serving as indicators of their workload.

The overall pipeline of root cause localization using monitoring KPI in Cloud $\mathcal{H}$ is shown in Fig.{\ref{framework}}. Cloud service providers typically have many data centers spread across different regions. Each region consists of multiple isolated locations known as availability zones to ensure low latency and high availability~\cite{kaushik2021cloud}. Users can create their VMs in any region that best fits their needs. Then, the behavior of both the host CPU cluster and the VMs is continuously monitored and recorded through KPIs, including CPU usage, memory usage, and netflow throughput. Next, KPI correlation analysis is conducted to understand the dependencies between each VM and the host cluster. Based on the KPI correlation analysis, mitigation strategies such as VM migration or throttling are enacted to alleviate the system overload. In our paper, we focus on the third and most significant part, namely root cause analysis, and propose \name. 

\subsection{A Motivating Example}

In a cloud system, there exist intrinsic correlations between the KPIs of individual VMs and the alarm KPI~\cite{gu2024identifying}, which is a crucial part of RCA. Take the CPU usage in cloud systems as an example, the correlation is based on the fundamental principle of resource allocation within a cloud system that each VM is allocated a portion of the cluster's resources like CPU~\cite{cortez2017resource}. When a VM's workload increases, it consumes more CPU resources, thereby affecting the overall CPU usage. However, the relationship between the KPIs of individual VMs and the overall CPU usage of the cluster is complex and non-linear~\cite{wang2023hierarchical}. This complexity is due to the sophisticated architecture of modern cloud systems and the principles of resource allocation they employ. In other words, these mechanisms ensure that the resource usage of one VM does not significantly impact others, thereby preventing a single VM from monopolizing the CPU~\cite{zhou2013scheduler}. Thus, the bulge of the workload KPI of a single VM does not necessarily lead to alarm KPI trigger alerts. 

To effectively identify the root cause of performance anomaly, we capture the correlations between the VM KPIs and the alarm KPI that depicts the contribution of VMs to the detected performance anomaly. This correlation often manifests in a similar waveform between the VM's KPIs and the alarm KPI. For example, a sudden surge in a VM's data traffic would likely lead to an increased demand for CPU resources, which would be reflected as a spike in the KPI of the cluster's CPU usage~\cite{beloglazov2012optimal}. The KPI correlation analysis approach aiming to mine the inherent correlations in KPI data can be leveraged to pinpoint the root causes of system alerts. In our case, similarity and causality analysis are adopted. Firstly, similarity analysis allows us to identify which VMs are behaving similarly to the overall system's performance, as reflected by the alarm KPI. Therefore, similarity analysis can help narrow down the potential root causes of the anomaly. Secondly, causality analysis is critical as it allows us to determine which changes in VM KPIs occurred before the anomaly, thus providing clues as to which VMs might have triggered the anomaly.

\begin{figure*}
\centering
\includegraphics[width=\linewidth]{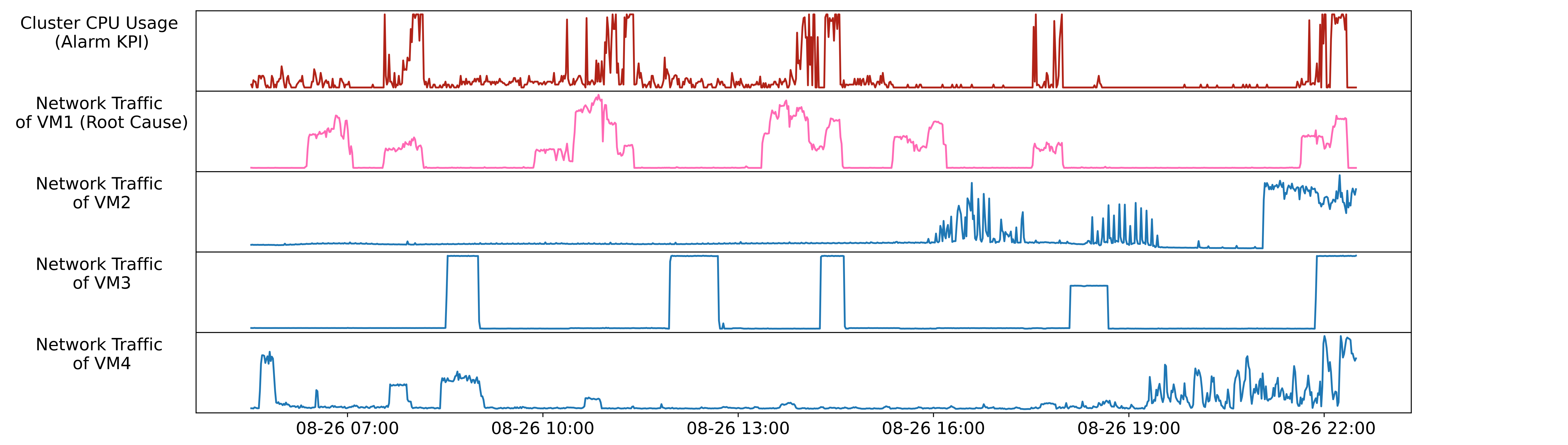}
\vspace{-4pt}
\caption{An Industrial Case in Cloud $\mathcal{H}$}
\vspace{-8pt}
\label{example}
\end{figure*}

An industrial case in a real-world cloud system cluster of Cloud $\mathcal{H}$ is shown in Fig.\ref{example}. There is an alarm KPI monitoring the overall CPU usage of the cluster, and several VM KPIs monitor the network traffic of individual VMs. For the purpose of the discussion, we will focus on four of the VM KPIs. We can observe that the waveforms of VM2 and VM4 have weak alignments with the fluctuations in the alarm KPI, indicating a lower correlation and, thus, are unlikely to be significant contributors to CPU overload. The KPI of VM1 and VM3 exhibit a high degree of similarity to the alarm KPI, indicating they are potential root causes for the anomaly. However, to ascertain the true root cause of the CPU overload, time series causality, \ie chronological order of events, should also be taken into consideration. As confirmed by the SREs, it is VM1, not VM3, which is the true root cause of the CPU overload. This is because the spike in VM1's KPI precedes the CPU overload anomaly, while the spike in VM3's KPI happens slightly after the anomaly, indicating that it is an outcome, not a cause of the anomaly. Indeed, in a cloud system, a VM's increase in resource consumption usually precedes the CPU overload due to temporal causality, which is why we take temporal causality into consideration in our method.

\subsection{\blue{Different Types of Performance Anomalies}}

\begin{figure*}
\centering
\includegraphics[width=.6\linewidth]{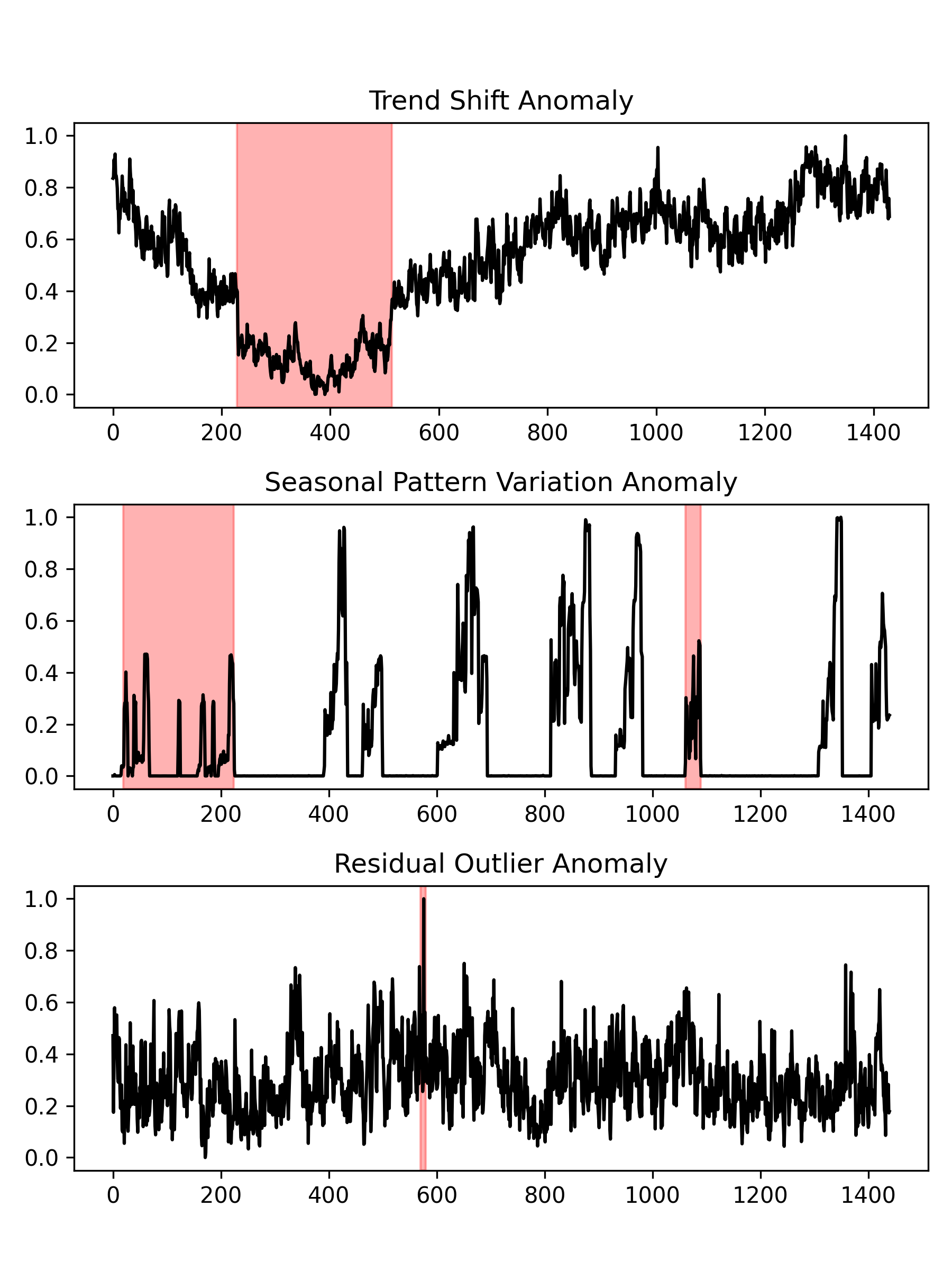}
\vspace{-1mm}
\caption{Different Anomaly Types in Cloud $\mathcal{H}$}
\vspace{-2mm}
\label{types}
\end{figure*}

\blue{Our previous work KPIRoot~\cite{gu2024kpiroot} predominantly focuses on detecting trend anomalies using a threshold-based method. While effective for identifying gradual or sustained shifts in performance metrics, this approach may not adequately capture the breadth of anomalies that can occur in Cloud $\mathcal{H}$. Specifically, seasonal and residual anomalies, which manifest as periodic deviations or abrupt, unexpected changes, respectively, might not be sufficiently detected by a threshold method alone.}

\blue{In Figure~\ref{types}, we observe three distinct types of performance anomalies across different monitoring metrics within Cloud $\mathcal{H}$. The first anomaly is a trend anomaly characterized by a sudden downward shift in throughput on a network interface card (NIC). This abrupt change can be indicative of packet loss, which might occur due to network congestion, hardware malfunctions, or configuration errors. The second case illustrates seasonal anomalies in NIC throughput, with unexpected deviations occurring in the area marked with red spans. The anomaly could suggest issues like batch jobs running at non-standard times or misconfigured scheduling that leads to throughput drop. The third example presents a residual anomaly in the average throughput on another NIC. Such short-duration spikes are neither part of a long-term trend nor follow a seasonal pattern, hinting at sporadic issues such as brief network outages, hardware failures, or security incidents like DDoS attacks. All these three types of performance anomalies can have severe impacts on service performance and reliability.}

\section{METHODOLOGY}

\blue{In this section, we present \name, an automated approach for root cause localization with monitoring KPIs in cloud systems. We first formulate the problem we target. Then, we provide an overview of the proposed method. Next, we elaborate on each part of our method, \ie time series decomposition-based anomaly segment detection, similarity analysis, and causality analysis. We finally analyze the complexity of our proposed algorithm.}

\subsection{Problem Formulation}

\begin{figure*}
\centering
\includegraphics[width=\linewidth]{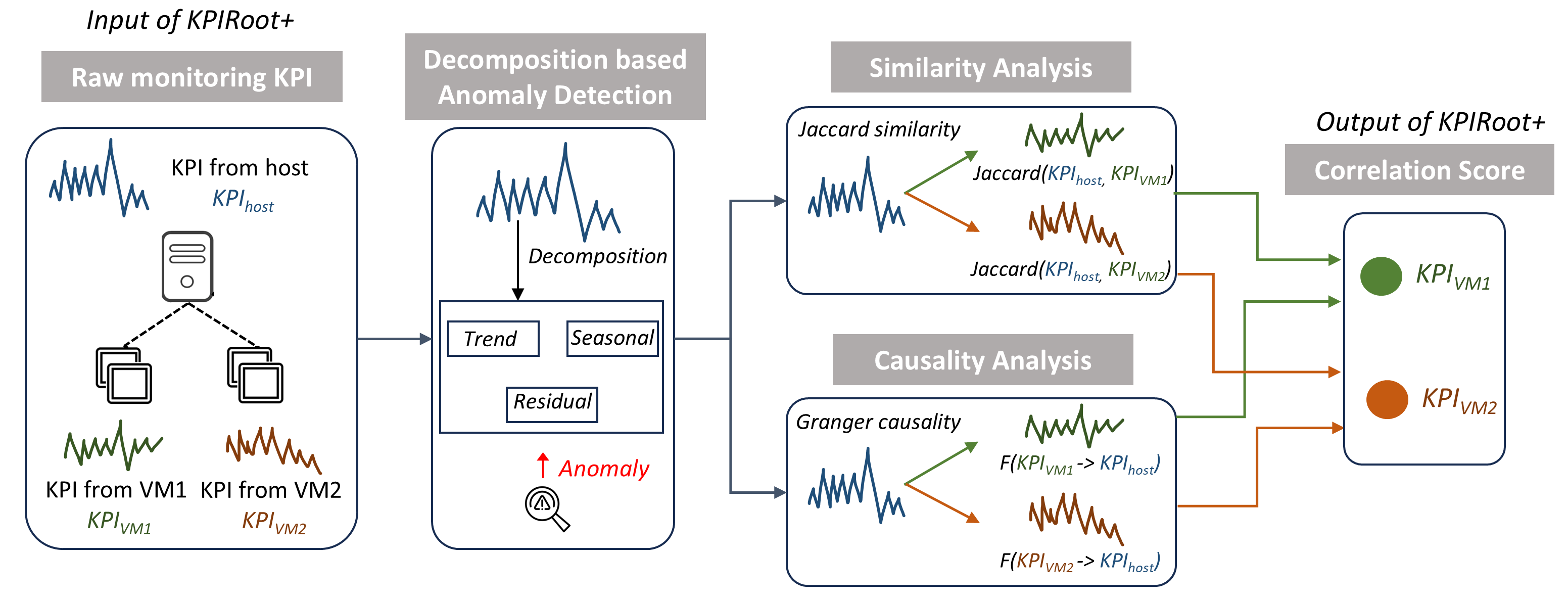}
\vspace{0mm}
\caption{The Overview of Our Proposed Method \name}
\vspace{0mm}
\label{overview}
\end{figure*}

The goal of our work is to identify the root causes of performance anomalies, including but not limited to CPU overload in large-scale cloud systems based on the alarm KPI and observed individual KPIs. The root causes are the VMs that influence the system service quality. By throttling the throughput of these VMs, we can alleviate the system-level anomaly and restore service quality. Given the alarm KPI that monitors the status of the host cluster $X_{host}\in{R^{n}}$ and the monitored KPIs of VMs, \eg the netflow of them $X_i\in{R^{n}, i\in\{1, 2, ..., m\}}$, where $N$ denotes the number of observations collected at an equal interval and $m$ is the number of monitored VMs. To determine the true root cause of the detected anomaly, a correlation score $c_i\in[0,1]$ that represents the contribution of a VM KPI to the anomaly is calculated. Then, the root causes can be obtained by ranking the correlation score, and KPIs with the top $K$ scores are deemed as root causes. 

\subsection{Overview}

\blue{The overview of \name is shown in Fig.\ref{overview}, which consists of three key components, namely, time series decomposition-based anomaly segment detection, similarity analysis, and causality analysis. Given the raw monitoring KPI, to make the RCA more efficient and meet the real-time requirement of industrial deployment, we propose to adopt SAX representation to downsample the raw KPI. Then, KPIRoot detects the potential anomaly segments, including different anomaly types in the downsampled alarm KPI of the host cluster (Section.\ref{anomaly}). In this step, an anomaly score that describes the probability of KPI being anomalous will be computed, an anomaly segment will automatically extracted around the spike. Then, KPIRoot conducts a similarity analysis to compute the similarity between VM KPIs and the alarm KPI during the anomaly period (Section.\ref{similarity}). This analysis provides insights into how each VM influences the host cluster by measuring the alignment of the KPI trends.  A causality analysis is then conducted (Section.\ref{causality}) to identify the cause-and-effect between the VM KPIs and the alarm KPI. In our case, we utilize Granger causality. The results from the similarity and causality analyses are then combined to compute a correlation score for each KPI.} 

\subsection{\blue{Time Series Decomposition Based Anomaly Segment Detection}}
\label{anomaly}

\blue{To make \name efficient and meet the industrial requirement of real-time identification, KPIRoot~\cite{gu2024kpiroot} propose to adopt Symbolic Aggregate Approximation (SAX)~\cite{lin2003symbolic}. SAX has several advantages in KPI analysis: First, SAX allows for a significant reduction in the dimension of the raw KPI, which can make subsequent similarity computation more efficient~\cite{minnen2007detecting}. Second, SAX can effectively filter out the noise and highlight the significant patterns in the KPIs by aggregating several consecutive data points into a single "symbol"~\cite{senin2013sax}. Specifically, the raw KPI $x$ of length $n$ will be represented as a $w$-dimensional vector $P=\{p_1, p_2,...,p_w\}$, where the $j^{th}$ element can be calculated as follows:}

\begin{figure}
\centering
\includegraphics[width=.8\linewidth]{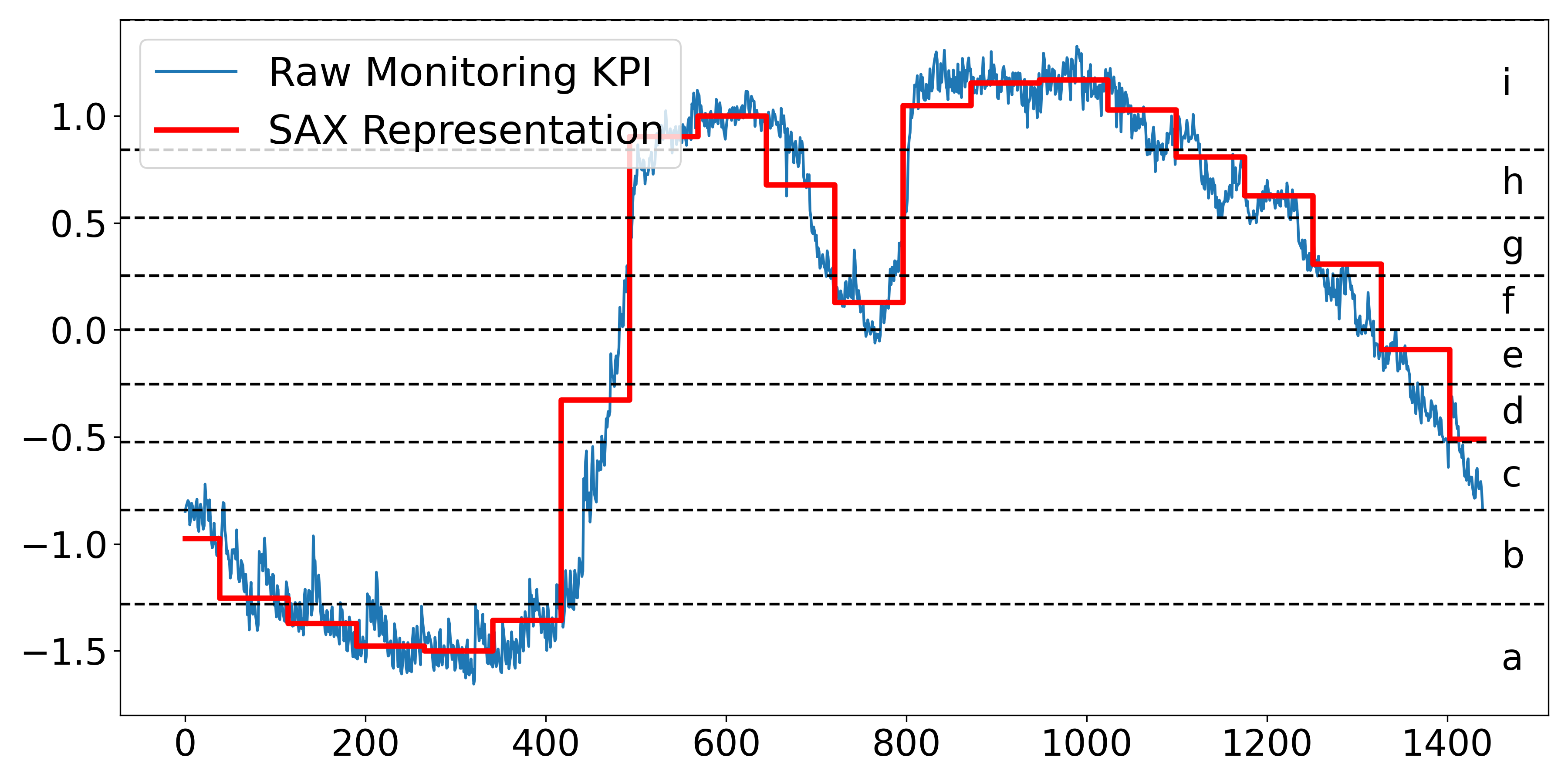}
\vspace{-1mm}
\caption{An Illustration of SAX Representation}
\vspace{-1mm}
\label{sax}
\end{figure}

\vspace{-4mm}
\begin{align}
    p_i = \frac{w}{n}\sum_{j=\frac{n}{w}(i-1)+1}^{\frac{n}{w}i}{x_j}
\end{align}
\vspace{-2mm}

\blue{In other words, to reduce the dimension of KPI from $n$ to $w$, the KPI is divided into $w$ equal-sized subsequences. The mean value of the subsequence is calculated, and a vector of these values becomes the Piecewise Aggregate Approximation (PAA) representation~\cite{guo2010improved}. Indeed, PAA representation is intuitive and simple yet shows an approximate performance compared with more sophisticated dimension reduction representations like Fourier transforms and wavelet transforms~\cite{lin2003symbolic}. Before converting it to the PAA, we normalize each KPI to have a mean of zero and a standard deviation of one. However, SAX representation can obscure significant variation trends due to its reliance on segment averages, potentially leading to inaccurate representations.}

\blue{In the industrial scenario, a fixed threshold method (e.g., CPU usage higher than 80\%) is commonly used to detect system resource usage anomalies. However, fixed thresholds can be limiting as they do not adapt to changes in the system's behavior over time. Typically, an anomaly refers to a state where the system's resources, such as CPU, memory, or network bandwidth, are being utilized at their maximum capacity and will cause performance issues for the system. However, in a dynamic cloud system, the threshold at which an anomaly occurs can shift. Specifically, during periods of low demand, a sudden spike in resource usage might be considered an anomaly. However, during peak demand periods, the system might be designed to handle much higher resource usage. Thus, the same usage level would not be considered an anomaly. Furthermore, the individual preferences of engineers make the setting of universally acceptable static thresholds complex. What might be a suitable threshold for one engineer could be too high or too low for another, leading to potential issues being overlooked or an excessive number of false alarms~\cite{zhao2020understanding}. KPIRoot assumes that by detecting an uprush in workload, the early warning of potential system anomaly can be identified, and root cause localization will be enabled. A score that describes the variation trend of a KPI is computed as follows:}  

\vspace{-4mm}
\begin{align}
    r_i = \frac{\sum_{k=i}^{i+l-1}p_k}{\sum_{j=i-l}^{i-1}p_j} 
\end{align}
\vspace{-2mm}

\noindent where $l$ denotes the historical lags taken into consideration. If the value $r_i$ is greater than a large threshold $\gamma$, it suggests that the usage of resources as indicated by the KPI starts to undergo a spike, and we denote the start point of overload as $t_s$. Once the KPI value drops below the value of $t_s$, it signifies that the overload ends; the endpoint of the overload is denoted as $t_e$. In other words, $x_{t_e}<x_{t_s}$ and $x_{t_e-1}>x_{t_s}$. 

\blue{However, KPIRoot~\cite{gu2024kpiroot} primarily targets trend anomalies through threshold-based techniques, which may fall short in identifying performance anomalies in large-scale cloud systems. The complexity and scale can lead to multiple overlapping types of performance anomalies, including level shifts, periodic variations, and sudden spikes or dips. We extend our previous work by proposing to utilize time series decomposition to better differentiate and detect these diverse anomaly types. This method distinctly identifies and addresses performance anomalies, which can often be obscured in a unified analysis.} 

\blue{We assume the metric time series can be decomposed as the sum of three different components, namely, trend, seasonality, and remainder components:}

\vspace{-2mm}
\begin{align}
    X_{host}^t = \tau_{host}^t+s_{host}^t+r_{host}^t, t=1,2,...,n
\end{align}
\vspace{-2mm}

\noindent \blue{where $X_{host}^t$ denotes the original host cluster KPI at time $t$, $\tau_{host}^t$ denotes the trend, $s_{host}^t$ denotes the periodic component and $r_{host}^t$ is the residual component.}

\blue{In this paper, we propose to use the Seasonal and Trend decomposition using the Loess (STL) algorithm, which is a robust and versatile method for decomposing time series data~\cite{rb1990stl}. It uses a sequence of Loess (locally estimated scatter plot smoothing) regressions. The flexibility of STL in handling various seasonal patterns and the ability to adjust its parameters makes it particularly suitable for complex and non-linear metrics in large-scale cloud systems.} 

\blue{After obtaining the decomposition into seasonal, trend, and remainder components, we perform anomaly detection on each component separately to identify distinct types of anomalies. To encode the complex patterns of the time series, it is necessary to consider both the local and global information, \ie multi-scale features. We adopt an auto-encoder network architecture with skip connections, as known as the U-Net structure~\cite{ronneberger2015u}. It is trained on multiple sliding window segments of the monitoring metrics. Although the autoencoder approach may incur some additional computational cost, it remains affordable, considering there is typically only one alarm KPI against thousands of VM KPIs.}

\subsection{Similarity Analysis}
\label{similarity}

\begin{figure}
\centering
\includegraphics[width=.8\linewidth]{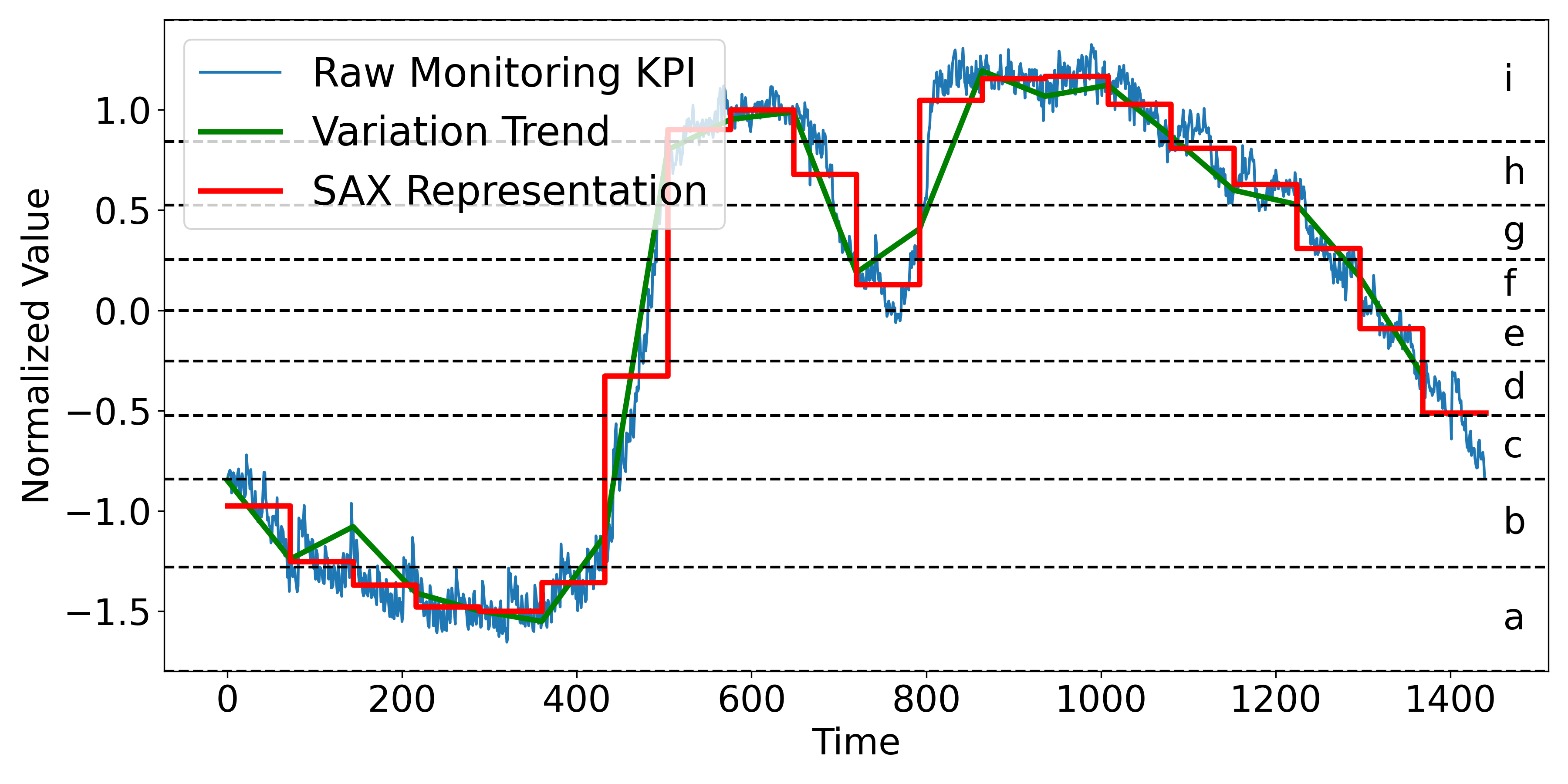}
\vspace{-1mm}
\caption{An Illustration of Improved SAX Representation}
\vspace{-1mm}
\label{isax}
\end{figure}

Motivated by~\cite{yang2021aid}, we propose to compute the similarity of the alarm KPI and VM KPIs to measure the degree of the root cause. The intuition behind this is that if a VM is responsible for triggering an overload, its KPI should exhibit a significant similarity with the host cluster's KPI, especially during periods of overload. If a VM is indeed the root cause of an overload, it is expected that its resource usage pattern would reflect the pattern of the host resource usage. 

Although there exist some approaches that can be used to calculate the similarity of monitoring KPIs, such as AID~\cite{yang2021aid}, HALO~\cite{zhang2021halo}, and CMMD~\cite{yan2022cmmd}, however, in real-time cloud computing systems, timely root cause localization is paramount. Traditional algorithms such as Dynamic Time Warping (DTW) might not be suitable for such scenarios due to their high time complexity, which can be prohibitive for processing large volumes of data in a real-time manner.

\blue{KPIRoot transforms the KPIs into symbolic sequences and then computes the similarity between these sequences using the Jaccard similarity coefficient. A discretization technique that produces symbols with equal probability is used to obtain the discrete representation with symbols. As proved by~\cite{lin2003symbolic}, the normalized KPIs have nearly Gaussian distributions. It’s easy to pick equal-sized areas under the Gaussian distribution curve using lookup tables for the cut line coordinates, slicing the under-the-Gaussian-curve area. Suppose there $\alpha$ symbols in the SAX representation, then the breakpoints refer to a sort of numbers $\beta=\{\beta_1,\beta_2,...,\beta_\alpha\}$ such that the area under normalized Gaussian distribution curve between $\beta_i$ to $\beta_{i+1}$ is equal to $\frac{1}{\alpha}$. The PAA representation element in Section.\ref{anomaly} between $\beta_i$ to $\beta_{i+1}$ will be assigned with the $i^{th}$ symbol shown as follows:}

\vspace{-4mm}
\begin{align}
    s_i = alphabet_l,\quad if\ {\beta_l}\leq{p_i}\leq\beta_{l+1} 
\end{align}
\vspace{-2mm}

\noindent where, $alphabet_i$ denotes the $i^{th}$ symbol and $s_i$ denotes the $i^{th}$ element of the SAX representation $S$. An example of SAX representation of a monitoring KPI with $w=20, \alpha=9$ is shown in Fig.\ref{sax}.

\blue{However, the traditional Symbolic Aggregate Approximation (SAX) method, while effective for dimensionality reduction and pattern recognition in time series data, has limitations due to its reliance on segment averages. This approach can obscure significant trends and variations, leading to misleading representations. For instance, two segments with different behaviors—such as an increasing CPU usage in a VM with relatively low average usage and a decreasing usage in another VM with a higher average, might be mapped to the same symbol if their averages are similar. This could overlook critical issues like potential CPU overloads. To address this, an Improved SAX representation (ISAX) is proposed (shown in Figure~\ref{isax}), which incorporates the variation trend indicators. To maintain the efficiency of the approach, only trend information is considered, ensuring that the enhanced representation remains computationally feasible while providing more critical insights into the KPI's dynamic behavior. The trend information, represented by the sign of the slope within the dimensionality reduction window, is calculated as follows:}

\vspace{-4mm}
\begin{align}
    \phi_i = sgn(x_{\frac{n}{w}\cdot{i}}-x_{\frac{n}{w}\cdot(i-1)+1})
\end{align}
\vspace{-2mm}

\noindent \blue{where $x_{\frac{n}{w}\cdot{i}}$ and $x_{\frac{n}{w}\cdot(i-1)+1}$ are the start and end point of the $i^{th}$ metric segment in PAA representation. The Improved SAX in \name will further differentiate between two metric segments that originally map to the same symbol under traditional SAX due to similar averages. By incorporating the variation trend, Improved SAX assigns different symbols to segments that have the same average but different trends, such as an increasing versus a decreasing sequence. Different from the original SAX representation, improved SAX will assign the PAA representation element between $\beta_i$ to $\beta_{i+1}$ a symbol as follows:}

\vspace{-4mm}
\begin{align}
    s_i = alphabet_{2{\alpha}-{\phi_i}\cdot{l}},\quad if\ {\beta_l}\leq{p_i}\leq\beta_{l+1} 
\end{align}
\vspace{-2mm}

We adopt the Jaccard similarity coefficient rather than other similarity measures because of its advantages when dealing with symbolic sequences like the SAX representation~\cite{he2016non}. Moreover, Jaccard similarity is easy to compute and can effectively capture the similarity between two symbolic sequences regardless of their lengths. This makes it very suitable for our case, where the lengths of the symbolic sequences could vary. Then, the Jaccard similarity can be computed as follows: 

\vspace{-4mm}
\begin{align}
    Jaccard(S_{host}, S_i) = \frac{\lvert{S_{host}\cap{S_i}}\rvert}{\lvert{S_{host}\cup{S_i}}\rvert} 
\end{align}
\vspace{-2mm}

\noindent where $S_{host}$ is the SAX representation of the host cluster's KPI and $S_i$ is the SAX representation of individual VM KPI $X_i$.

\subsection{Causality Analysis}
\label{causality}

\blue{The Improved Symbolic Aggregate Approximation method is effective in reducing the dimension of raw KPI while preventing trend information loss; however, the computation of Improved SAX representation-based similarity does not provide any insights into the causality between VM KPIs and alarm KPIs. As mentioned by~\cite{mariani2020predicting}, the ability of Granger causality analysis to analyze the correlation between KPIs can be a key factor for improving the accuracy of the root cause localization. By using Granger Causality in conjunction with SAX representation, we can not only analyze large quantities of time series data effectively but also gain insights into the potential causality between different KPIs. That is why we take Granger Causality~\cite{shojaie2022granger} as a supplement.} 

Granger Causality is a statistical hypothesis test used to determine if one KPI is useful in forecasting another KPI~\cite{arnold2007temporal}. For instance, if a VM KPI undergoes an uprush and causes the alarm KPI to trigger alerts, \ie the change in the VM KPI precedes the changes in the alarm KPI, then Granger causality exists from the alarm KPI to the VM KPI. It should be noted that Granger Causality is unidirectional, which means that if VM KPI Granger causes alarm KPI, it does not imply that alarm KPI Granger causes VM KPI. In our case, we are interested in understanding how VM KPIs influence the alarm KPI of the host cluster, so we focus on the Granger causality from the VM KPIs to the alarm KPI. Specifically, assuming that the two KPIs can be well described by Gaussian autoregressive processes, the autoregression (AR) of alarm KPI without and with information from VM KPI can be written as follows:

\vspace{-4mm}
\begin{align}
    p_{alarm}^{t} = \hat{a_0} + \sum_{j=1}^{q}{\hat{a_j}}p_{alarm}^{t-j} + \hat{\varepsilon_t}
\end{align}
\vspace{-2mm}

\vspace{-4mm}
\begin{align}
    p_{alarm}^{t} = a_0 + \sum_{j=1}^{q}{a_j}p_{alarm}^{t-j} + \sum_{j=1}^{q}{b_j}p_{i}^{t-j} + \varepsilon_t
\end{align}
\vspace{-2mm}

\noindent where the first equation uses the past values of the PAA representation of host KPI $X^{host}$ while the second includes the past values of the PAA representation of both $X^{host}$ and $X^{vm}$. Furthermore, $\hat{a_j}$ is the autoregression coefficients for $X^{host}$, while $a_j$ and $b_j$ are the autoregression coefficients for $X^{host}$ with the contribution of both $X^{host}$ and $X^{vm}$'s historical values. Both $\hat{\varepsilon_t}$ and $\varepsilon_t$ are residual terms assumed to be Gaussian, and $q$ is model order, which represents the amount of past information that will be included in the prediction of the future sample. Then, we conduct the F-statistic test:

\vspace{-4mm}
\begin{align}
    F_{vm\rightarrow{host}} = \frac{\sum_{t=t_s+q}^{t_e}({\hat\varepsilon_t^2}-{\varepsilon_t^2})/q}{\sum_{t=t_s+q}^{t_e}{\varepsilon_t^2}/(t_{e}-t_{s} - 2q - 1)}
\end{align}
\vspace{-2mm}

\noindent where ${\hat\varepsilon_t^2}$ and $\varepsilon_t^2$ represent the mean square error (MSE) of the AR model of host KPI without and with information from VM KPI. $t_s$ and $t_e$ are the start point and end point of the detected overload. The F-statistic test follows an F-distribution with $q$ and $t_e-t_s - 2p - 1$ degrees of freedom under the null hypothesis that the VM KPI does not Granger-cause the host KPI. The calculated F-statistic can be a good indicator of the VM KPI Granger-causality to the host KPI.

After both the similarity and causality analyses are performed, KPIRoot combines these two scores to create a more comprehensive correlation score for each VM KPI. Specifically, the correlation score is a weighted sum of similarity score and causality score:

\vspace{-4mm}
\begin{align}
    c_i = \lambda\times{Jaccard(S_{host}, S_i)} + (1-\lambda)\times{F_{vm\rightarrow{host}}}
\end{align}
\vspace{-2mm}

\noindent where $c_i$ is the correlation score between the $i^{th}$ VM KPI and the alarm KPI. The balance weight $\lambda$ is a hyper-parameter. In our experiments, this parameter is set to be 0.9.

\subsection{Complexity Analysis}
\label{complexity}

The proposed method \name is summarized in Algorithm.\ref{algo:1}. The computation of our method mainly lies in the similarity and causality analysis. In industrial practice, $w\approx{\sqrt{n}}$, which means the lengths of SAX representation of KPIs are roughly $\sqrt{n}$. So, the time complexity of obtaining SAX representation is $\mathcal{O}(\sqrt{n})$. On one hand, the time complexity of Jaccard similarity is directly proportional to the KPI length, so the complexity of similarity analysis is $\mathcal{O}(\sqrt{n})$. On the other hand, the complexity of Granger causality mainly depends on the autoregression of $P_{host}$, which is $\mathcal{O}(\sqrt{n}\times{q^3})$, where $q$ is the time lag of Granger causality (usually very small). Thus, the complexity of KPIRoot is $\mathcal{O}(\sqrt{n}\times({q^3}+2))$. As a comparison, the time efficiency of methods like AID (based on DTW) is $\mathcal{O}(n^2)$, let alone more complex deep learning-based methods like CMMD. Therefore, \name is a more suitable method for industrial applications that demand real-time root cause localization. 

\begin{algorithm}[t]
    \SetAlgoLined
    \caption{KPI Root Cause Localization+}
    \label{algo:1}
    \normalsize
    \begin{algorithmic}[1] 
        \REQUIRE The alarm KPI of the host $X_{alarm}$; The KPIs of VMs $X_i, i\in\{1, 2, ..., m\}$; 
        \ENSURE The correlation scores of VM KPIs that correlate to the anomaly of alarm KPI $c_i$
        \FOR {$i = 1 $; $ i \leq w $; $ i ++ $}
        \STATE {$p^i_{alarm} = \frac{w}{n}\sum_{j=\frac{n}{w}(i-1)+1}^{\frac{n}{w}i}{x_{alarm}^j}$}
        \STATE {$\phi_{alarm}^i = sgn(x_{alarm}^{\frac{n}{w}\cdot{i}}-x_{alarm}^{\frac{n}{w}\cdot(i-1)+1})$}
        \ENDFOR
        \STATE // Anomaly Segment Detection
        \STATE {$X_{alarm}^t = \tau_{host}^t+s_{host}^t+r_{host}^t$}
        \STATE {$i_{anomaly}=AE(\tau_{host}){\cup}AE(s_{host}){\cup}AE(r_{host})$}
        \STATE {$p_{alarm} = p_{alarm}[i_{anomaly}]$}
        \STATE {$s^i_{alarm} = \{alphabet_{2{\alpha}-{\phi_i}\cdot{l}},\quad if\ {\beta_l}\leq{p^i_{alarm}}\leq\beta_{l+1}\}$}
        \FOR {$ i = 1 $; $ i \leq m $; $ i ++ $}
        \STATE // Similarity Analysis
        \FOR {$ k = 1 $; $ k < m $; $ k ++ $}
        \STATE {$p_i^{k} = \frac{w}{n}\sum_{j=\frac{n}{w}(k-1)+1}^{\frac{n}{w}k}{x_i^k}$}
        \STATE {$p_i = p_i[i_{anomaly}]$}
        \STATE {$\phi_{i}^k = sgn(x_{i}^{\frac{n}{w}\cdot{k}}-x_{i}^{\frac{n}{w}\cdot(k-1)+1})$}
        \STATE {$s_i^{k} = \{alphabet_{2{\alpha}-{\phi_i}\cdot{l}},\,s.t.\ {\beta_l}\leq{p_i^k}\leq\beta_{l+1}$\}}
        \ENDFOR
        \STATE {$Jaccard(S_{host}, S_i) = \frac{\lvert{S_{host}\cap{S_i}}\rvert}{\lvert{S_{host}\cup{S_i}}\rvert}$}
        \STATE // Causality Analysis
        \FOR {$ t = t_s + q $; $ t < t_e $; $ t ++ $}
        \STATE {$p_{alarm}^{t} = \hat{a_0} + \sum_{j=1}^{q}{\hat{a_j}}p_{alarm}^{t-j} + \hat{\varepsilon_t}$}
        \STATE {$p_{alarm}^{t} = a_0 + \sum_{j=1}^{q}{a_j}p_{alarm}^{t-j} + \sum_{j=1}^{q}{b_j}p_{i}^{t-j} + \varepsilon_t$}
        \ENDFOR
        \STATE {$F_{vm\rightarrow{host}} = \frac{\sum_{t=t_s+q}^{t_e}({\hat\varepsilon_t^2}-{\varepsilon_t^2})/q}{\sum_{t=t_s+q}^{t_e}{\varepsilon_t^2}/(t_{e}-t_{s} - 2q - 1)}$}
        \STATE {$c_i = \lambda\times{Jaccard(S_{host}, S_i)} + (1-\lambda)\times{F_{vm\rightarrow{host}}}$}
        \ENDFOR
        \RETURN {$c_i$}
    \end{algorithmic} 
\end{algorithm}
\setlength\textfloatsep{4mm}

\section{EVALUATION}

\blue{To fully evaluate the effectiveness of our proposed approach, \name, we use three real-world monitoring KPI datasets from the cloud service systems of Cloud $\mathcal{H}$. Particularly, we aim to answer the following research questions (RQs):}

\begin{itemize}[leftmargin=*]
    \item \blue{RQ1: How effective is \name in performance issue detection compared with baselines?}
    \item {RQ2: How effective is \name compared with KPI root cause localization baselines?}
    \item {RQ3: How effective is each component of \name in root cause localization?}
    \item {RQ4: How efficient is \name in localizing root cause KPIs compared to baselines?}
    \item \blue{RQ5: How sensitive is \name to each hyperparameter?}
\end{itemize}

\begin{table}[t]
\centering
\vspace{0pt}
\caption{Statistics of Industrial Dataset}
\begin{tabular}{cccc}
\toprule
Industrial & Dataset A & Dataset B & Dataset C\\  
\midrule
Host Clusters & 16 & 6 & 7\\
\midrule
VM Number & 120$\sim$803 & 21$\sim$26 & 41$\sim$57\\
\midrule
KPI Length & 5,928,480 & 17,040 & 37,200\\
\midrule
Root Causes & 4$\sim$36 & 3$\sim$8 & 2$\sim$15\\
\bottomrule
\end{tabular}
\vspace{-0pt}
\label{table1}
\end{table}

\subsection{Experiment Setting}

\subsubsection{Datasets}

To confirm the practical significance of KPIRoot, we collect three datasets from large-scale online services in three Available Zones (AZs) of Cloud $\mathcal{H}$. The statistics of three industrial datasets are shown in Table~\ref{table1}. Various VM KPIs and alarm KPIs monitor the status of the service. The VM KPIs typically measure the healthy status of each VM, including resource usage metrics like CPU, memory, I/O, and bandwidth usage. The alarm KPI monitors the runtime status at the host cluster level, which is usually positively correlated to the VM KPIs. 

\subsubsection{Evaluation Metrics}

In the following experiments, the F1-score is utilized to evaluate the performance of root cause localization results. We employ Precision: $PC=\frac{TP}{TP+FP}$, Recall: $RC=\frac{TP}{TP+FN}$, F1 score: $F1=2\cdot\frac{PC\cdot{RC}}{PC+RC}$. To be specific, $TP$ is the number of correctly localized VM KPIs; $FP$ is the number of incorrectly predicted VM KPIs; $FN$ is the number of root cause VM KPIs that failed to be predicted by the model. F1 score is the harmonic mean of the precision and recall. In real-world applications, since the number of root cause KPIs is unknown, software engineers will first investigate top $k$ recommended results by root cause localization methods. Hit Rate@$k$ is a widely used metric to measure whether the correct root causes (in our case, the root cause VM KPIs) are within the recommended top $k$ results. We adopt Hit Rate@$5$ and Hit Rate@$10$ as evaluation metrics in our experiments. 

\blue{Additionally, we propose to use the Normalized Discounted Cumulative Gain (NDCG) in our evaluation metrics, specifically NDCG@$10$. NDCG is more beneficial because it considers the rank position of each result, applying a discounting factor to lower-ranked positions, which measures how easily engineers can find the culprit VMs. This is crucial in our scenarios as the most relevant root causes are more prioritized for investigation. NDCG@$1$ is left out because it is the same as Hit Rate@$1$ in our scenario. NDCG@$k$ measures to what extent the root cause appears higher up in the ranked candidate list. Thus, the higher the above measurements, the better.}

\subsection{Experimental Results}

\begin{table*}[htbp]
\centering
\caption{Experimental Results of Different Anomaly Detection Methods}
\scalebox{0.9}{\begin{tabular}{c|ccc|ccc|ccc}
\toprule
\multirow{2}{*}{Methods} & \multicolumn{3}{c|}{Dataset A} & \multicolumn{3}{c|}{Dataset B} & \multicolumn{3}{c}{Dataset C}\\  
 & Pre & Rec & F1 & Pre & Rec & F1 & Pre & Rec & F1\\ 
\midrule
3$\sigma$ & 0.709 & 0.762 & 0.738 & 0.765 & 0.694 & 0.730 & 0.797 & 0.685 & 0.747\\
LOF & 0.681 & 0.587 & 0.753 & 0.619 & 0.591 & 0.737 & 0.681 & 0.598 & 0.715\\
IF & 0.699 & 0.612 & 0.788 & 0.673 & 0.607 & 0.772 & 0.715 & 0.612 & 0.706\\
Autoencoder & 0.791 & 0.770 & 0.782 & 0.776 & 0.810 & 0.794 & 0.859 & 0.793 & 0.823\\
LSTM & 0.836 & 0.752 & 0.805 & 0.826 & 0.865 & 0.824 & 0.829 & 0.863 & 0.839\\
KPIRoot & 0.787 & 0.712 & 0.755 & 0.782 & 0.717 & 0.744 & 0.803 & 0.709 & 0.759\\
\midrule
\name & \textbf{0.914} & \textbf{0.943} & \textbf{0.928} & \textbf{0.924} & \textbf{0.907} & \textbf{0.913} & \textbf{0.942} & \textbf{0.863} & \textbf{0.894}\\
\bottomrule
\end{tabular}}
\vspace{-4pt}
\label{anomaly}
\end{table*}

\subsubsection{\blue{\textbf{RQ1} The effectiveness of \name in performance issue detection}}

\blue{To answer this research question, we compare the performance of \name with five widely used performance anomaly detection methods in cloud systems, 3$\sigma$~\cite{yu2023nezha}, LOF (Local Outlier Factor)~\cite{breunig2000lof}, IF (Isolation Forest)~\cite{liu2008isolation}, Autoencoder~\cite{xu2018unsupervised}, LSTM~\cite{zhao2021predicting} and KPIRoot~\cite{gu2024kpiroot}. The results are shown in Table~\ref{anomaly}, where the best Precision, Recall and F1 scores are all marked with boldface. We can see that the average Precision, Recall and F1 scores of \name outperform all baseline methods in three datasets, including our previous method, KPIRoot. Each of the baseline methods has its strengths depending on the specific type of anomaly. Methods like $3\sigma$, LOF, and IF are particularly effective at detecting residual anomalies because they are point-wise anomaly detection, which identifies deviations from normal behavior at specific data points. This makes them suitable for catching sudden or isolated anomalies but less effective for detecting anomalies that persist over time. On the other hand, Autoencoder and LSTM models are designed to capture deviations from historical patterns by embedding the sliding window of metrics and fitting local patterns. These methods are effective at identifying seasonal anomalies, where recurring deviations from the periodic patterns happen. KPIRoot, in contrast, computes the variation between the current observation window and previous observation windows, which makes it particularly adept at identifying trend anomalies. This method is well-suited for detecting gradual changes or level shifts in performance metrics over time.} 

\blue{Despite the capabilities of these individual approaches, they often yield suboptimal results when all anomaly types are mixed together, as they cannot differentiate between them effectively. This is where \name demonstrates its superiority by utilizing a time series decomposition-based method. By decomposing time series data into its constituent components, \name is able to better isolate and identify trends, seasonal patterns, and residual anomalies, leading to higher accuracy and more comprehensive anomaly detection. Furthermore, the effective identification of performance anomalies is crucial not only for immediate anomaly detection but also for facilitating the subsequent root cause localization.}

\subsubsection{\textbf{RQ2} The effectiveness of \name}

\blue{To answer this research question, we compare the performance of \name with several other methods, including three statistical correlation measurements: Kendall correlation, Spearman correlation, and CloudScout~\cite{yin2016cloudscout}. Additionally, we consider AID~\cite{yang2021aid}, which uses DTW distance, LOUD~\cite{mariani2018localizing}, a graph centrality-based method, HALO~\cite{zhang2021halo}, which employs conditional entropy, CMMD~\cite{yan2022cmmd}, a graph neural network-based method, and our previously proposed method, KPIRoot~\cite{gu2024kpiroot}. Table~\ref{table2} presents the results, highlighting the best F1 scores, Hit@5, Hit@10, and NDGG@10 in bold. We observe that \name consistently outperforms all baseline methods across three datasets in terms of average F1 scores, Hit@5, Hit@10, and NDGG@10. In particular, the improvement achieved by \name is more pronounced in Dataset B and Dataset C compared to Dataset A. This is because these datasets focus on KPIs, such as request rates, related to the load balancer, which manages the distribution of network traffic across physical machines. As a result, anomalies in VM request rates tend to precede anomalies in host clusters, providing an early indicator for potential issues. It is important to note that, as shown in Table~\ref{table2}, the number of root causes often exceeds 5. Consequently, not all root causes can be captured within the top 5 predictions. Despite this, achieving Hit@5 scores exceeding 70\% is significant, as it indicates that our method accurately identifies a substantial portion of root causes within just the top 5 predictions. Additionally, the high F1 score and Hit@10 demonstrate the method's effectiveness for industrial applications.} 

\blue{We can observe that baseline models like Kendall, Spearman, CloudScout, and AID have worse performance. These coefficient-based methods fundamentally measure the similarity between the shape of KPIs. However, high similarity does not necessarily imply causality because a high similarity can occur due to a shared underlying cause rather than one KPI directly influencing another KPI. Though CMMD has the ability to capture complex, nonlinear relationships between KPIs through graph attention neural networks and achieves a Hit@10 of 0.801$\sim$0.848, it still falls short of considering the causality between VM KPIs and the host cluster KPI. HALO computes the conditional entropy between VM KPIs and the host KPI, which somehow alleviates the defect of neglecting the causality between KPIs. The LOUD method applies graph centrality to pinpoint the root causes of issues. However, the way in which the graph is constructed can significantly impact the results. As a result, the LOUD method fails to deliver optimal performance, making it less effective in accurately identifying the root causes of problems in our context. KPIRoot incorporates both the similarity analysis through SAX representation similarity and causality analysis through the Granger causality test, leading to better root cause localization accuracy than other baselines. Compared with KPIRoot, \name can identify performance anomalies more accurately and comprehensively, thereby enhancing the accuracy of subsequent root cause analysis. The improved SAX technique utilized by \name helps retain trend variation information, thereby reducing false positives and enhancing the overall robustness of anomaly detection.}

\begin{table*}[htbp]
\centering
\caption{Experimental Results of Different Root Cause Localization Methods}
\scalebox{.73}{\begin{tabular}{c|cccc|cccc|cccc}
\toprule
\multirow{2}{*}{Methods} & \multicolumn{4}{c|}{Dataset A} & \multicolumn{4}{c|}{Dataset B} & \multicolumn{4}{c}{Dataset C}\\  
 & F1 & H@5 & H@10 & N@10 & F1 & H@5 & H@10 & N@10 & F1 & H@5 & H@10 & N@10\\ 
\midrule
Kendall & 0.651 & 0.562 & 0.728 & 0.507 & 0.605 & 0.594 & 0.770 & 0.546 & 0.657 & 0.635 & 0.727 & 0.651\\
Spearman & 0.681 & 0.587 & 0.753 & 0.518 & 0.619 & 0.591 & 0.737 & 0.577 & 0.681 & 0.598 & 0.715 & 0.636\\
CloudScout & 0.699 & 0.612 & 0.788 & 0.657 & 0.673 & 0.607 & 0.772 & 0.683 & 0.715 & 0.612 & 0.706 & 0.608\\
LOUD & 0.736 & 0.652 & 0.813 & 0.624 & 0.736 & 0.625 & 0.824 & 0.657 & 0.709 & 0.653 & 0.829 & 0.689\\
AID & 0.746 & 0.652 & 0.749 & 0.634 & 0.673 & 0.618 & 0.794 & 0.602 & 0.665 & 0.613 & 0.729 & 0.597\\
HALO & 0.734 & 0.651 & 0.842 & 0.667 & 0.632 & 0.569 & 0.811 & 0.598 & 0.719 & 0.635 & 0.789 & 0.646\\
CMMD & 0.776 & 0.632 & 0.833 & 0.604 & 0.679 & 0.594 & 0.848 & 0.613 & 0.721 & 0.667 & 0.801 & 0.658\\
KPIRoot & 0.859 & 0.731 & 0.909 & 0.766 & 0.860 & 0.749 & 0.946 & 0.779 & 0.829 & 0.713 & 0.895 & 0.741\\
\midrule
\blue{\name} & \blue{\textbf{0.884}} & \blue{\textbf{0.780}} & \blue{\textbf{0.934}} & \blue{\textbf{0.823}} & \blue{\textbf{0.891}} & \blue{\textbf{0.799}} & \blue{\textbf{0.967}} & \blue{\textbf{0.842}} & \blue{\textbf{0.871}} & \blue{\textbf{0.755}} & \blue{\textbf{0.936}} & \blue{\textbf{0.797}}\\
\bottomrule
\end{tabular}}
\vspace{-4pt}
\label{table2}
\end{table*}

\begin{table*}[htbp]
\centering
\caption{Experimental Results of the Ablation Study of \name}
\scalebox{.7}{\begin{tabular}{c|cccc|cccc|cccc}
\toprule
\multirow{2}{*}{Methods} & \multicolumn{4}{c|}{Dataset A} & \multicolumn{4}{c|}{Dataset B} & \multicolumn{4}{c}{Dataset C}\\  
 & F1 & H@5 & H@10 & N@10 & F1 & H@5 & H@10 & N@10 & F1 & H@5 & H@10 & N@10\\ 
\midrule
KPIRoot \textit{w/o} I & 0.872 & 0.763 & 0.935 & 0.789 & 0.883 & 0.766 & 0.963 & 0.793
& 0.856 & 0.740 & 0.922 & 0.784\\
KPIRoot \textit{w/o} D & 0.865 & 0.752 & 0.926 & 0.770 & 0.872 & 0.762 & 0.958 & 0.784 & 0.845 & 0.734 & 0.909 & 0.766\\
\midrule
\blue{\name} & \blue{\textbf{0.884}} & \blue{\textbf{0.780}} & \blue{\textbf{0.934}} & \blue{\textbf{0.823}} & \blue{\textbf{0.891}} & \blue{\textbf{0.799}} & \blue{\textbf{0.967}} & \blue{\textbf{0.842}} & \blue{\textbf{0.871}} & \blue{\textbf{0.755}} & \blue{\textbf{0.936}} & \blue{\textbf{0.797}}\\
\bottomrule
\end{tabular}}
\vspace{-6pt}
\label{table3}
\end{table*}

\subsubsection{\textbf{RQ3} The effectiveness of components in \name}

\blue{To answer this research question, we conducted an ablation study on \name. We compared two baseline models, removing the Improved SAX and Decomposition-based anomaly detection part of \name to investigate the contribution of these two designs.}

\begin{itemize}[leftmargin=*]
    \item \blue{{\textit{\name w/o I} This baseline removes the Improved SAX and utilizes the SAX representation in KPIRoot. The Decomposition-based anomaly detection is adopted.}}
    \item \blue{{\textit{\name w/o D} This baseline removes the Decomposition-based anomaly detection and utilizes the Improved SAX representation to downsample the original metrics.}}
\end{itemize}

\blue{Table~\ref{table3} shows the performance comparison between \name and its variants. In summary, the effectiveness of \name is enhanced with the utilization of Improved SAX and Decomposition-based anomaly detection. Indeed, the variant without the Improved SAX performs better than the variant without Decomposition-based anomaly detection. This is because accurate anomaly detection is crucial for conducting subsequent similarity and causality analyses, which are essential for correlating the true root cause. While the trend information captured during downsampling by Improved SAX is also important, as the increasing or decreasing trend within a downsample window can sometimes be crucial for determining the true root cause, there are subtle differences between the variation trends of the true root cause and false positives. Both variants outperform the original KPIRoot, demonstrating that the integration of these two designs significantly boosts the root cause localization performance.}

\begin{figure}
\centering
\includegraphics[width=\linewidth]{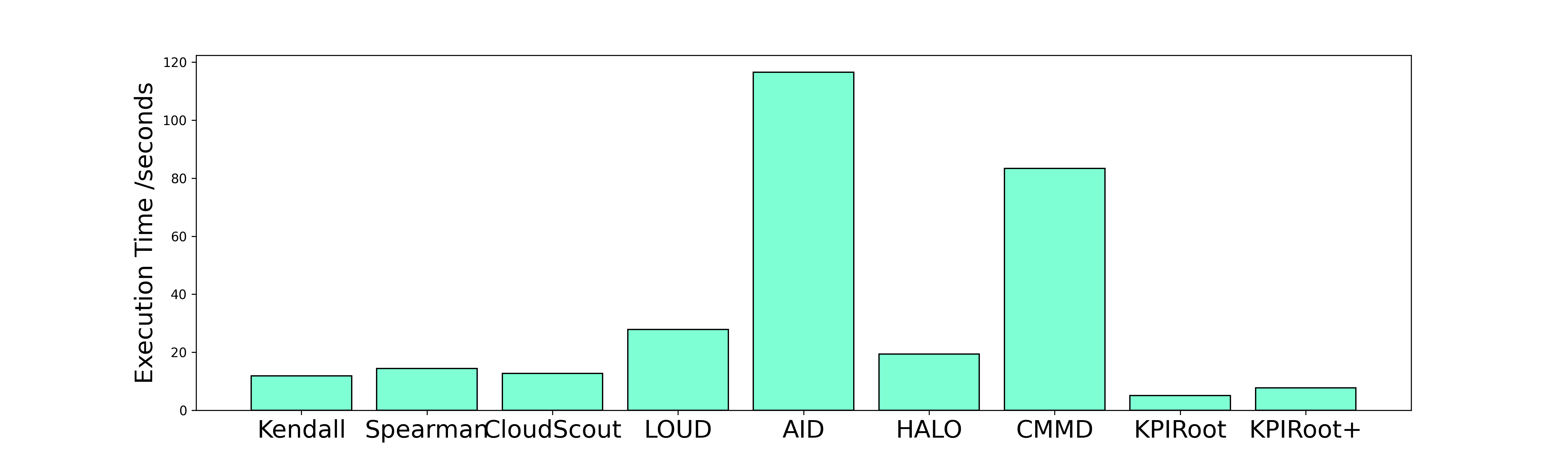}
\vspace{-8pt}
\caption{Root Cause Localization Time for All Methods}
\vspace{-4pt}
\label{efficiency}
\end{figure}

\subsubsection{\textbf{RQ4} The efficiency of \name}

\blue{In this section, we evaluate the efficiency of \name in large-scale cloud systems of Cloud $\mathcal{H}$. The average running time of each method is shown in Fig.~\ref{efficiency}, from which we can observe that KPIRoot is still the most efficient, with an average execution time of around only 5 seconds.  While \name takes around 8 seconds, it is still capable of providing real-time analysis and delivers more accurate results, making it a worthwhile tradeoff. The additional overhead is primarily due to the decomposition-based anomaly detection method and the Improved SAX, which uses more symbols, making the similarity analysis more time-consuming. However, this overhead is absolutely acceptable, given the improved accuracy and comprehensiveness of the results. This indicates that KPIRoot is capable of providing real-time root cause analysis, meeting the requirements of large-scale cloud systems where timely identification of root causes is critical. As for methods like AID and CMMD, their performances are less than satisfactory due to their inherent computational complexities. AID, with its time complexity of $\mathcal{O}(n^2)$, suffers from an average runtime of more than one hundred seconds. On the other hand, CMMD, which applies graph attention neural networks, requires high computational resources, which also leads to a slower execution time and makes it less efficient. Therefore, both AID and CMMD fail to deliver the desired levels of efficiency, particularly in large-scale, real-time environments. Baseline methods like Kendall and Spearman may seem appealing due to their lower computation times. However, these apparent gains are offset by their inferior accuracy levels. As a result, their use can lead to inaccurate root-cause diagnoses and, subsequently, ineffective problem-solving solutions.}

\blue{In summary, the evaluation results highlight \name’s superior accuracy while not adding much more computational overhead, thereby offering an excellent balance between efficiency and precision in real-time root cause analysis. It is a highly promising tool for conducting real-time root cause analysis within large-scale cloud systems.}

\subsubsection{\blue{\textbf{RQ5} Sensitivity Analysis of \name}}

\begin{figure}
  \centering
  \subfigure[Parameter Sensitivity of $w$]{
  \label{Sensitivity1}
  \includegraphics[width=0.44\columnwidth]{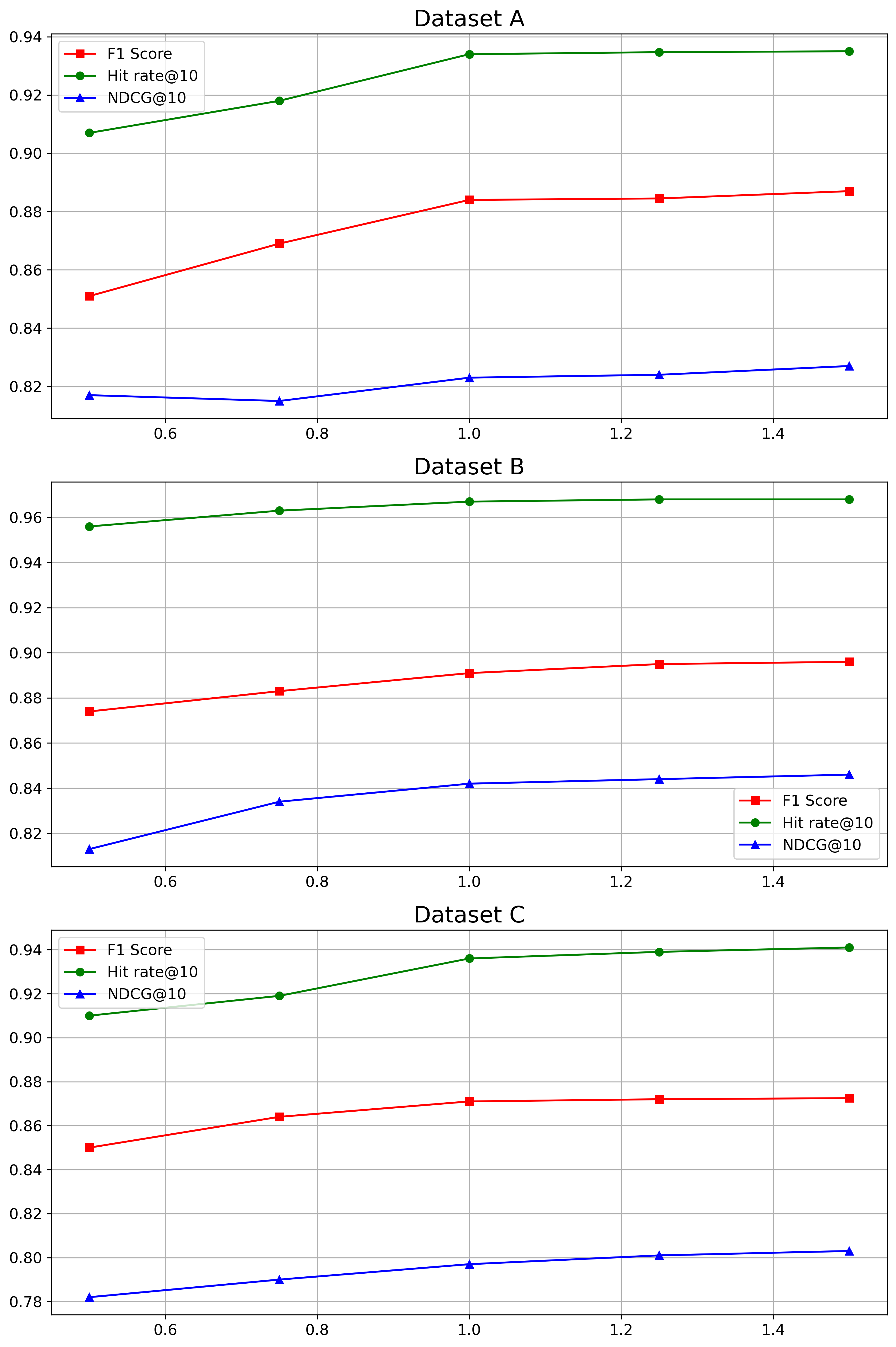}
  }
  \subfigure[Parameter Sensitivity of $\lambda$]{
  \label{Sensitivity2}
  \includegraphics[width=0.44\columnwidth]{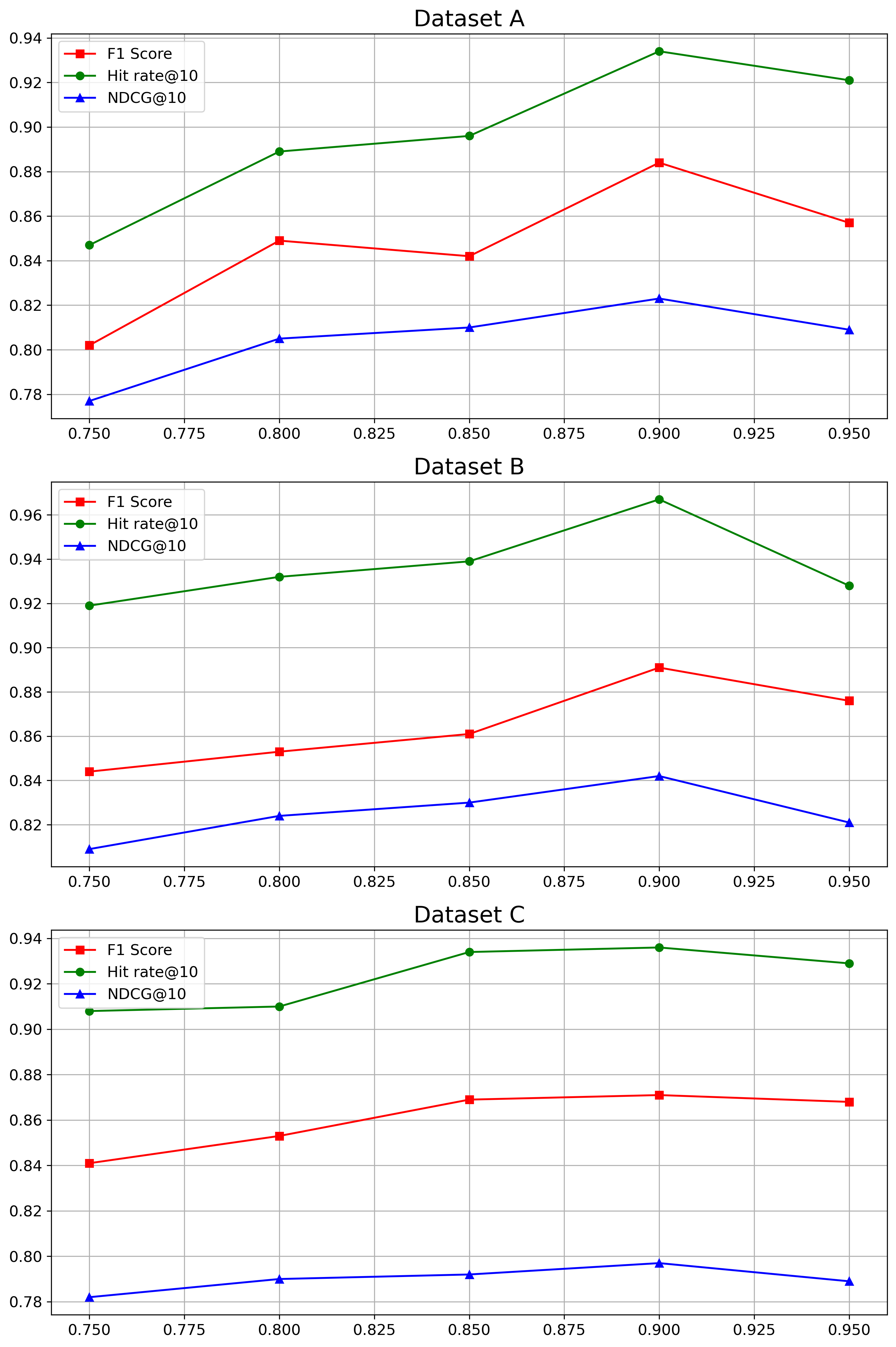}
  }
  \vspace{-4pt}
  \caption{Parameter Sensitivity of \name
  }
  \vspace{-4pt}
  \label{Sensitivity}
\end{figure}

\blue{The parameter $w$ determines the dimension of the representation vector for Improved SAX, while $\lambda$ is crucial for balancing the tradeoff between similarity and causality analysis. We evaluate the sensitivity of \name to these two hyper-parameters using three industrial datasets. To ensure fairness, we vary the values of $w$ and $\lambda$ while keeping all other parameters constant. Specifically, $w$ is chosen as a multiple of $\sqrt{n}$, ranging from 0.5 to 1.5 times $\sqrt{n}$. This selection ensures that we maintain an $O(\sqrt{n})$ time complexity, aligning with our efficiency goals. For $\lambda$, we select values from 0.75 to 0.95, acknowledging that the scale of similarity is typically smaller than that of causality. This choice effectively balances the tradeoff between similarity and causality within our analysis framework. By systematically adjusting these parameters, we aim to optimize the performance and robustness of our model across different datasets.}

\blue{Figure~\ref{Sensitivity} presents the experimental results of RQ5. For the parameter $w$, the performance is relatively stable between 1 to 1.5 times $\sqrt{n}$. If the dimension of Improved SAX is too low, there is more information loss during the downsampling process, which decreases accuracy. However, a larger dimension may cause the time complexity to increase quickly, and it may not significantly enhance performance beyond $\sqrt{n}$. Thus, it is reasonable to select $w$ in this range to balance the tradeoff between computational efficiency and model accuracy. For the parameter $\lambda$, a good tradeoff between these two parts indeed helps improve the performance of \name. In Dataset C, this variation of performance due to parameter is not so significant because either the similarity score or causality score of the root cause is high. Thus, the tradeoff coefficient has a lower influence on the overall performance.}

\begin{figure}
\centering
\includegraphics[width=.8\linewidth]{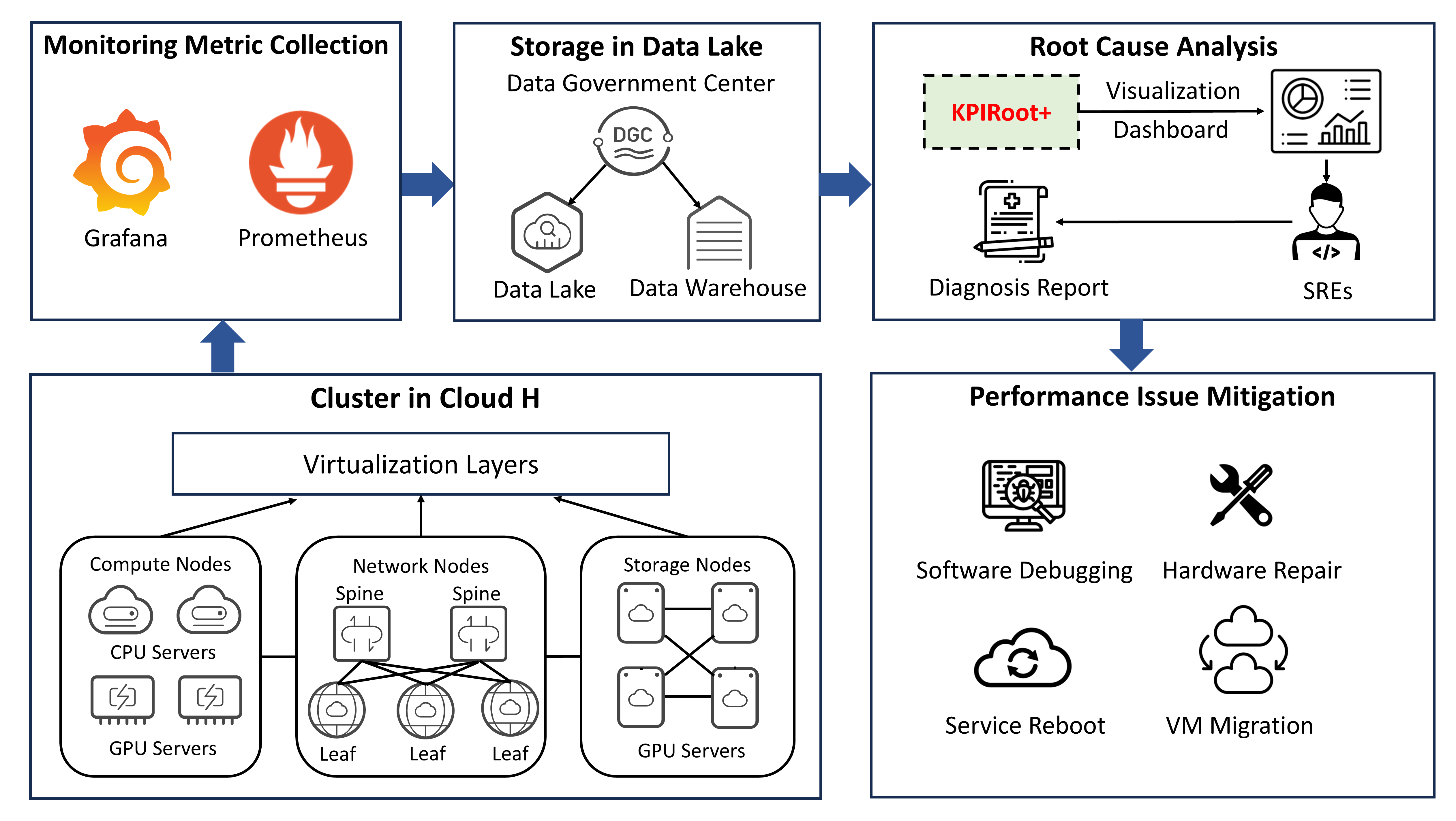}
\vspace{-2mm}
\caption{The overall pipeline of deploying \name in Cloud $\mathcal{H}$}
\vspace{-2mm}
\label{industry}
\end{figure}

\section{Industrial Experience}

In this section, we share our experience of deploying \name in the cloud system of Cloud $\mathcal{H}$, a full-stack cloud system that consists of an infrastructure layer, a platform layer, and an application layer. To support a large number of customers, each of our services is supported by multiple clusters with tens of hundreds of virtual instances (\eg virtual router) or devices. The collective workload of each cluster is continuously monitored using an alarm KPI. When abnormal traffic impacts these services, for instance, due to overwhelming requests overloading a service, an anomaly is swiftly detected based on the alarm KPI. This triggers a root cause analysis procedure to pinpoint the specific nodes (\eg, VMs) and take prompt mitigating actions. In our previous practice, manual inspection is feasible given the limited scale of each cluster. So, we can check each specific KPI of the node, compare it with the alarm KPI (with similarity comparison tools), and find the root cause. However, this process proved to be error-prone and labor-intensive, particularly as the scale of each service expanded. On average, it took between thirty minutes to one hour to identify and mitigate the root causes.

\begin{figure}
\centering
\includegraphics[width=\linewidth]{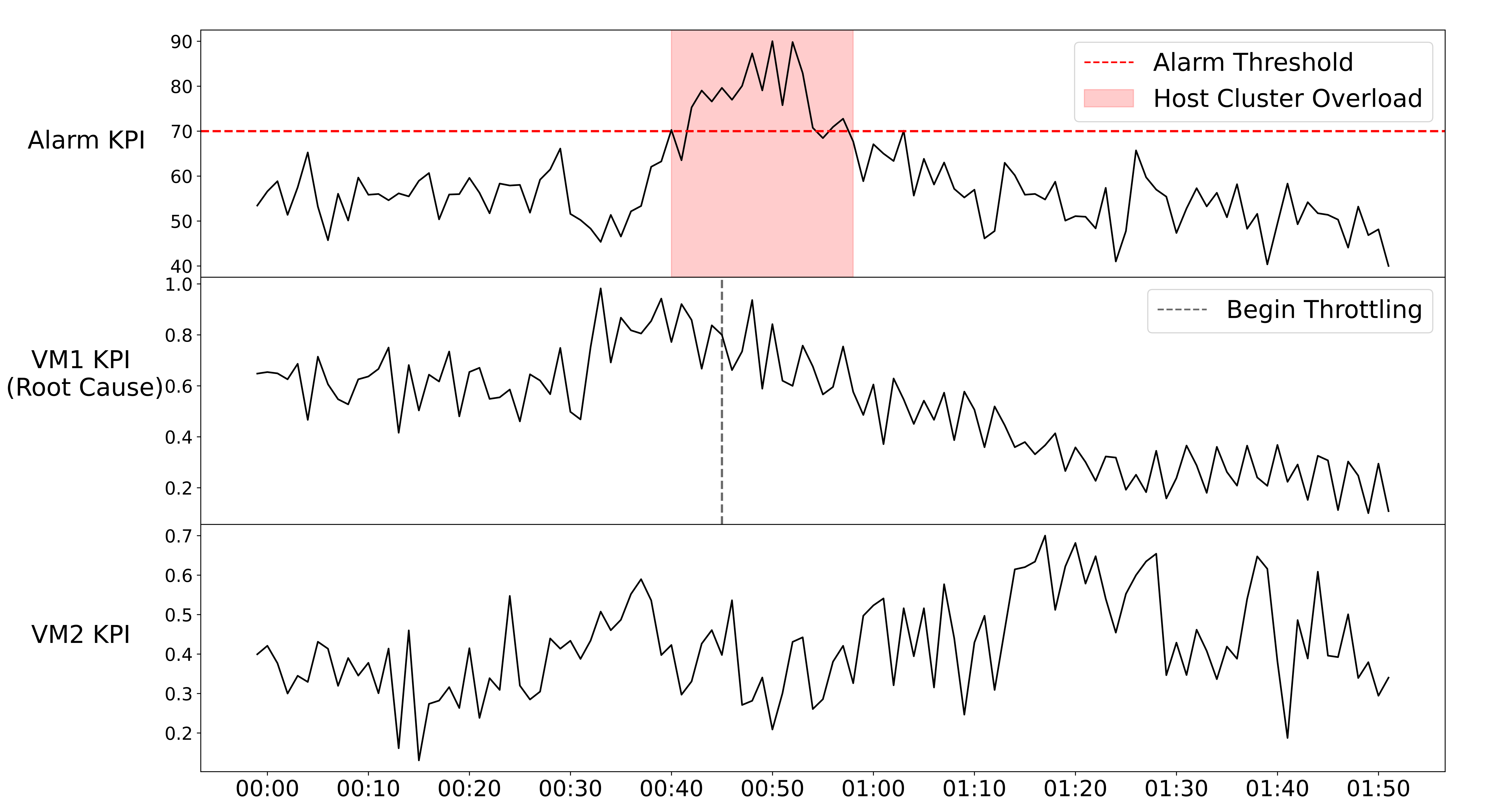}
\vspace{-2mm}
\caption{Case Study of KPIRoot}
\vspace{-1mm}
\label{case}
\end{figure}

\begin{figure}
\centering
\includegraphics[width=\linewidth]{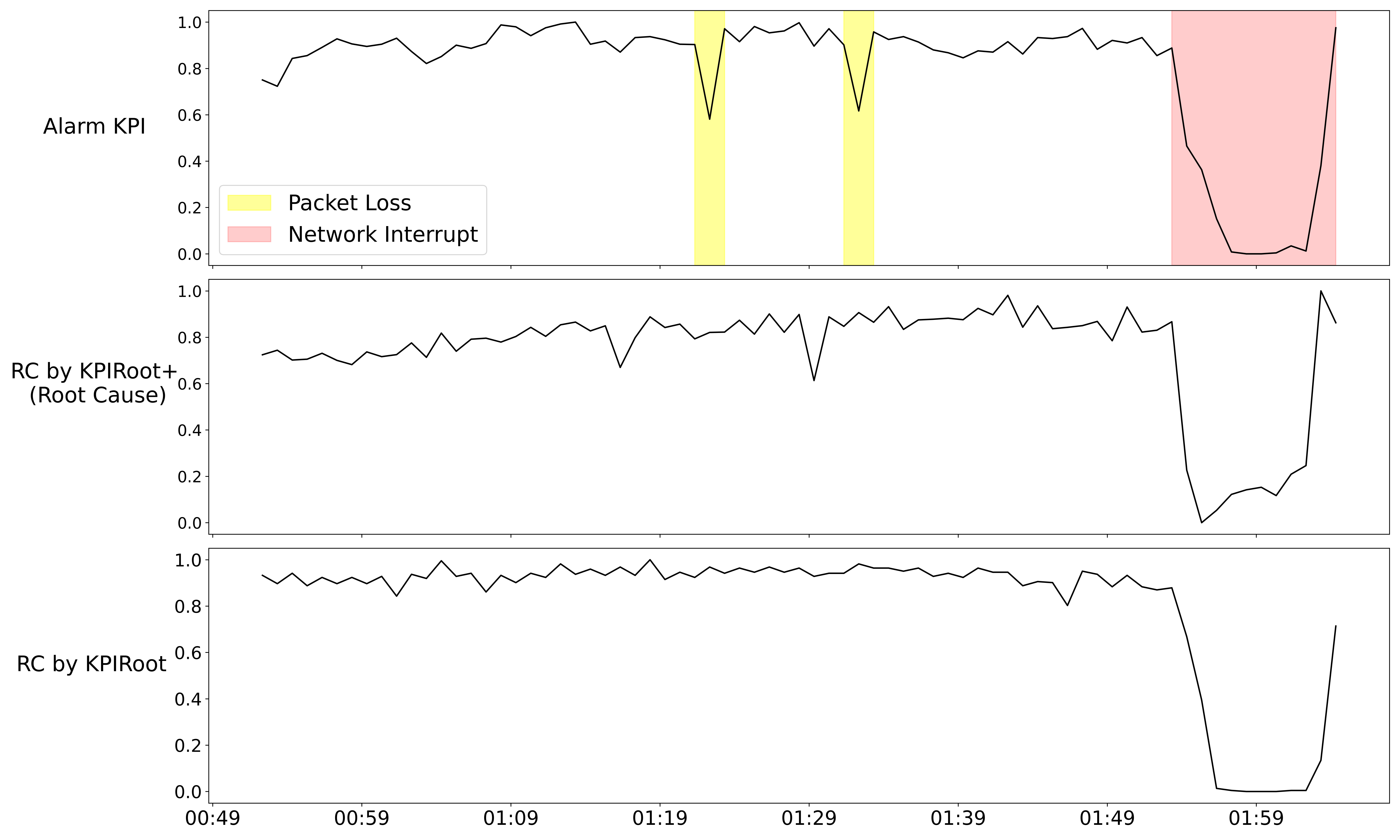}
\vspace{-2mm}
\caption{Case Study of KPIRoot}
\vspace{-1mm}
\label{new_case}
\end{figure}

\blue{To alleviate these issues, we have deployed \name in Cloud $\mathcal{H}$ since \textit{Aug 2023}. Specifically, \name operates by automatically fetching KPIs collected from the monitoring backends and applying the algorithm to calculate the correlation score in real time. Using KPIRoot, the potential root causes are returned to engineers. In addition, visualization tools are provided, making it easier for engineers to understand the system's behavior and performance. This overall deployment pipeline of deploying \name in Cloud $\mathcal{H}$ is depicted in Figure~\ref{industry}. The software reliability engineers collect the monitoring metrics (like CPU usage, network traffic, memory usage) of the host clusters and each VM through monitoring tools like Grafana, Prometheus, etc~\cite{agarwal2023outage}. Then, these monitoring metrics are stored in the Data Lake of Huawei Cloud, a highly scalable and flexible storage system that consists of the Data Lake Storage, the Data Warehouse, and the Data Lake Governance Center (DGC). Data Lake Storage is the actual storage space where all the data, including the monitoring metrics, are stored, while the Data Warehouse is an enterprise system used for reporting and data analysis. On top of these two components, the DGC is responsible for managing the data stored in the data lake and overseeing the lifecycle of the data, from ingestion and storage to usage and deletion. With Cloud $\mathcal{H}$'s data lake, real-time data analysis is enabled, \ie as soon as monitoring metrics are collected and stored in the data lake, they can be immediately accessed and analyzed by the performance issue diagnosis system empowered with \name. The results of \name are visualized on the dashboard and can be easily observed and understood by engineers. Once the root cause has been investigated and identified, the SREs prepare a diagnosis report, including a detailed description of the identified root cause and potential mitigation strategies.}

\blue{In Fig.~\ref{case}, the practical application of our previous root cause analysis tool, KPIRoot, in real industrial scenarios is shown. In this case, we initially received an alert indicating that the overall traffic for the host cluster had abruptly surpassed the predefined threshold. This requires immediate measures to pinpoint the root cause and throttle its throughput to avoid resource exhaustion within the cluster. However, this is quite challenging given the large number of KPIs needed to check, and the root-cause KPI may not be readily identifiable visually, as its shape similarity may not correspond directly with the alarm KPI. Given that, the root cause analysis takes tens of minutes to one hour to check manually, leading to delayed mitigation of the sudden traffic spike. With KPIRoot, the root cause of KPI can be quickly localized, generally within five minutes. With this result, we throttle the throughput of VM1 immediately after the alarm KPI is fired. As shown in Figure~\ref{case}, the overall traffic is limited, and the alarm KPI returns back to a normal range quickly. However, KPIRoot primarily considers trend anomalies and sometimes neglects critical performance anomaly information and recommends inaccurate root causes. A case that we identified is shown in Figure~\ref{new_case}. The frequent packet loss suggested that the load balancer was not distributing traffic evenly, causing the VM corresponding to the second KPI to experience frequent network congestion. As requests accumulated, it led to more severe network interruptions, which is symptomatic of overwhelming network buffers and potential misconfigurations in the load-balancing algorithm. This buildup of unsent packets can cause buffer overflow and increased latency, resulting in temporary network interruptions. Therefore, the second KPI was indeed the root cause, while the third KPI drop was a passive consequence and not the root cause. These two packet loss anomalies indicated by transient KPI drops in the alarm KPI would be ignored by KPIRoot. KPIRoot will instead recommend the third KPI as the root cause. In contrast, \name considers these types of anomalies, correctly identifying the second KPI as the root cause.} 

\name has been deployed in all major regions of our company, covering eighteen critical network services, \eg Linux Virtual Server (LVS), NGINX, Network Address Translation (NAT), and DNS services. It has been serving in our production environment for more than ten months, reducing the average root cause localization time from 30 minutes to 5 minutes. Following the deployment of the KPIRoot service, the feedback from engineers has been overwhelmingly positive. In terms of computational efficiency, KPIRoot has reduced the computational load significantly compared to previous methods. The system can perform real-time RCA, identifying potential issues quickly and allowing engineers to take immediate action. In terms of accuracy, KPIRoot's design of combining similarity and causality analysis has proven highly precise in identifying root causes. This leads to more effective problem resolution and significantly reduced revenue loss.

\section{Discussion}

In this section, we discuss the difference between our approach and existing root cause analysis approaches for microservice systems and why they are not applicable in our industrial scenario. We identified some potential threats to the validity of our study.

\subsection{Root Cause Analysis for Microservice System}

Our objective shares some similarities with root cause analysis in microservice systems; however, there are several main differences in terms of the application scenarios. Firstly, rather than localizing the root causes of application/service failures in microservice systems, where these applications are at the same level, our problem is top-down root localization. When we observe an anomaly at the system level, we investigate and analyze the underlying VM instance-level information. Secondly, due to VM isolation, each VM instance operates independently and is isolated from other VMs and the host system. This leads to sparse or even non-existent invocation dependency among them, making the construction of a service dependency graph as done in existing works very challenging.

Existing Methods like FRL-MFPG~\cite{chen2023frl} and ServiceRank~\cite{ma2021servicerank} rely on the construction of a service dependency graph and the execution of a second-order random walk, which can become highly time-consuming with complexity exceeding $O(n^2)$. As for HRLHF~\cite{wang2023root}, the large graph size makes causal discovery computationally intensive. Furthermore, the delay incurred by waiting for engineers to provide human feedback poses an additional obstacle for real-time localization. However, the analysis delay should be less than the sampling interval, \eg 1 minute in our practical scenarios, making these methods unsuitable for industrial deployment. 

\subsection{Threats to Validity}

We have identified the following potential threats to the validity of our study:

\textbf{Internal threats.} The implementation of baselines and parameter settings constitutes one of the internal threats to our work's validity. To mitigate these threats, we utilized the open-sourced code released by the authors of the papers or packages on GitHub for all baselines. As for our proposed approach, the source code has been reviewed meticulously by the authors, as well as several experienced software engineers, to minimize the risk of errors and increase the overall confidence in our results. For parameter settings, as our algorithm KPIRoot has few parameters, we find the most suitable configurations based on the best results obtained in different parameters.

\textbf{External threats.} 
Our experiments are conducted based on real-world datasets collected from Cloud $\mathcal{H}$ over more than two years. The evaluation requires engineers to inspect and label the root cause KPIs manually. Label noises are inevitable during the manual labeling process. However, alleviation strategies taken by engineers further ensure the accuracy of labeled root causes. Therefore, we believe the amount of noise is small and does not have a significant impact on the experiment results. On the other hand, the results may vary between different cloud service providers, industries, or specific use cases. Nevertheless, we believe that our experimental results, obtained from large-scale online systems within a prominent cloud service company serving millions of users, can demonstrate the generality and effectiveness of our proposed approach, KPIRoot.

\section{RELATED WORK}

\subsection{Anomaly Detection in Cloud Systems}
Ensuring the optimal performance of cloud systems is an imperative task. Monitoring KPIs are used to perceive the status of the cloud systems and facilitate analysis when performance anomalies occur. Many works~\cite{zong2018deep, chen2022adaptive, lin2018predicting, xu2018unsupervised, ren2019time, su2019robust} have been proposed for proactively discovering the unexpected or anomalous behaviors of the multivariate monitoring metric. Anomaly detection in cloud systems has been an important and widely studied topic as it ensures the reliability and efficiency of cloud systems. However, anomaly detection is regarded as a black box module that only predicts whether an anomaly happens, which is not enough for engineers to troubleshoot the system failure. In other words, once a performance anomaly has been detected in a cloud service system, further analyses should be enacted to pinpoint some abnormal metrics that are likely to be the possible root causes of that performance anomaly.

\subsection{Root Cause Localization in Cloud Systems}

Determining the root cause of performance anomalies for online service systems has been a hot topic. The goal of root cause localization with monitoring metrics data in cloud systems is to localize a subset of the monitored KPIs. Then, they can troubleshoot these specific parts of the system to alleviate the performance anomaly. LOUD~\cite{mariani2018localizing} assumes that the services of anomalous KPIs are likely to result in anomalous behavior of services it correlates with. Thus, LOUD applies graph centrality to identify the degree of the KPIs that correlate to the observed performance anomaly. AID~\cite{yang2021aid} is an approach that measures the intensity of dependencies between monitoring KPIs of cloud services. It calculates the similarities between the status KPIs of the caller and the callee. Then, AID aggregates the similarities to produce a unified value as the intensity of the dependency. It can also be deployed as a root cause localization tool as it can output the similarity between monitoring metrics and the KPI that triggers alerts. Similarly, CloudScout~\cite{yin2016cloudscout} employs the Pearson Correlation Coefficient over KPIs at the physical machine level, such as CPU usage, to calculate the similarity between services. 

There are also many works focusing on searching fault-indicating attribute combinations of KPI data. CMMD~\cite{yan2022cmmd} is proposed to perform cross-metric root cause localization through a graph attention network to model the relationship between fundamental and derived metrics. While HALO~\cite{zhang2021halo} proposed a hierarchical search approach to capture the relationship among attributes based on conditional entropy and locate the fault-indicating combination. Another approach iDice~\cite{lin2016idice} treats the root cause as a combination of attribute values, \ie the anomaly can be easily identified through the co-occurrence of some specific attribute dimensions. A Fisher distance-based score function is utilized for ranking the combination of the attributes, and effective combinations will be output. However, iDice is not suitable for large-scale issue reports with high-dimensional metrics from cloud systems. MID~\cite{gu2020efficient} employs a meta-heuristic search that automatically detects dynamic emerging issues from large-scale issue reports with higher efficiency. 

It is worth noting that, in our case, the monitoring metrics are not aggregated along different attribute dimensions through complex calculations of the raw data. Indeed, the monitoring metrics in our scenario directly reflect the run-time state of an entity, \eg the throughput of a client VM. In our practice, obtaining the root cause at a granularity of metric level is enough for engineers to troubleshoot the performance anomalies. Thus, we formulate our problem as localizing a subset of the monitored KPIs.

\section{CONCLUSION}

In this paper, we propose \name, an effective and efficient framework for anomaly detection and root cause analysis in practical cloud systems with monitoring KPIs. Specifically, \name is an improved version of KPIRoot that utilizes time decomposition-based anomaly detection and improved SAX representation, offering more accurate root cause localization results, while not compromising the efficiency. Extensive experiments on three industrial datasets show that KPIRoot achieves 0.882 F1-Score and 0.946 Hit@10 with the highest efficiency, outperforming all the baselines, including KPIRoot. Moreover, the successful deployment of our approach in large-scale industrial applications further demonstrates its practicality. 

\section*{COMPLIANCE WITH ETHICAL STANDARDS}

\textbf{Conflict of Interest} The authors have no competing interests to declare that are relevant to the content of this article.

\noindent\textbf{Funding} The work described in this paper was supported by the Research Grants Council of the Hong Kong Special Administrative Region, China (No. CUHK 14206921 of the General Research Fund) and Fundamental Research Funds for the Central Universities, Sun Yat-sen University (No. 76250-31610005).

\noindent\textbf{Ethical approval} This manuscript extends our previous ISSRE paper titled 'KPIRoot: Efficient Monitoring Metric-based Root Cause Localization in Large-scale Cloud Systems,' which further improves the KPIRoot. The authors also declare that this manuscript follows the best scientific standards, in particular with regard to acknowledgment of prior works, honesty of the presentation of results, and focus on the demonstrability of the statements. This manuscript and the work that led to it do not carry any specific ethical issue.

\noindent\textbf{Informed consent} All the authors give their consent to submit this work.

\noindent\textbf{Author Contributions} Data curation: Xinying Sun, Yongqiang Yang; Funding acquisition: Michael R. Lyu, Guangba Yu, Jiazhen Gu; Methodology: Wenwei Gu; Supervision: Michael R. Lyu; Validation: Renyi Zhong; Visualization: Jinyang Liu, Yintong Huo, Zhuangbin Chen, Jianping Zhang; Original Draft: Wenwei Gu; Review: Guangba Yu, Jiazhen Gu.

\noindent\textbf{Data Availability} The full data cannot be made available due to the privacy policy in Cloud $\mathcal{H}$. Only a portion of desensitized samples will be made public. The code is released in: {\href{https://github.com/WenweiGu/KPIRoot}{https://github.com/WenweiGu/KPIRoot}}.

\bibliography{emse25}

\end{document}